\documentclass{jfm}
\usepackage{graphicx}
\usepackage{amsmath}
\usepackage{epstopdf, epsfig}
\usepackage{color}
\usepackage{float}

\shorttitle{Translation of bubbles insonated with short acoustic pulses}
\shortauthor{E. Igualada-Villodre {\it et al.}}

\title{Transient effects in the translation of bubbles insonated with acoustic pulses of finite duration}

\author{Elena Igualada-Villodre\aff{1},
  Ana Medina-Palomo\aff{1},\\
  Patricia Vega-Mart\'{\i}nez\aff{1}
 \and Javier Rodr\'{\i}guez-Rodr\'{\i}guez\aff{1}\corresp{\email{javier.rodriguez@uc3m.es}}}

\affiliation{\aff{1}Fluid Mechanics Group, Carlos III University of Madrid,
Leganes, Madrid, 28911, SPAIN}

\newcommand{\rmd}{\mathrm{d}}
\renewcommand{\vec}{\mathbf}
\newcommand{\rme}{\mathrm{e}}
\newcommand{\erf}{\mathrm{erf}}
\newcommand{\erfc}{\mathrm{erfc}}
\newcommand{\Icos}{\mathrm{Icos}}
\newcommand{\Isin}{\mathrm{Isin}}
\newcommand{\F}{\mathrm{F}}
\newcommand{\Sto}{\mbox{\textit{St}}}
\newcommand{\We}{\mbox{\textit{We}}}
\newcommand{\Cd}{C_d}
\newcommand{\vvinf}{\vec{v}_{\infty}}
\newcommand{\vvinft}{\tilde{\vec{v}}_{\infty}}
\newcommand{\vUt}{\tilde{\vec{U}}}
\newcommand{\vWt}{\vec{\tilde{W}}}

\newcommand{\Wt}{{\tilde{W}}}

\newcommand{\dfr}[2]{\frac{\rmd #1}{\rmd #2}}
\newcommand{\dfi}[2]{\frac{\rmd\phantom{#2}}{\rmd #2}#1}

\begin{document}

\maketitle

\begin{abstract}
The translation of a bubble under the action of an acoustic forcing finds applications in fields ranging from drug delivery to sonoluminesce. This phenomenon has been widely studied for cases where the amplitude of the forcing remains constant over time. However, in many practical applications, the duration of the forcing is not long enough for the bubble to attain a constant translational velocity, mainly due to the effect of the history force. Here, we develop a formulation, valid in the limit of very viscous flow and small-amplitude acoustic forcing, that allows us to describe the transient dynamics of bubbles driven by acoustic pulses consisting of a finite number of cycles. We also present an asymptotic solution to this theory for the case of a finite-duration sinusoidal pressure pulse. This solution takes into account both the history integral term and the transient period that the bubble needs to achieve steady radial oscillations, being the former dominant during most of the acceleration process. Moreover, by introducing some additional assumptions, we derive a simplified formula that describes fairly well the time evolution of the bubble velocity. Using this solution we show that the convergence to the steady translational velocity, given by the so-called Bjerknes force, occurs rather slowly, namely as $\tau^{-1/2}$, which explains the slow convergence of the bubble velocity and stresses the importance of taking into account the history force.
\end{abstract}


\section{\label{sec:intro}Introduction}

In 1906, Bjerknes wrote ``A pulsating body in a synchronously oscillating current is subject to the action of a resultant force, the direction of which is that of the acceleration in the current at the time when the pulsating body has its maximum volume.'' \citep{Bjerknes1906}. The force described here, commonly known as primary Bjerknes force, results from the coupling between the volume oscillations of a body (a bubble in our case) and the pulsation of the external flow. Although the translational motion of a bubble in these circumstances is mainly oscillatory, over many oscillation cycles a net displacement exists, being this the manifestation of the force described by Bjerknes.

Nowadays, there exist different scientific problems and technological applications where the possibility of transporting bubbles via acoustic waves is of great relevance. One of the most promising usages of this force concerns the possibility of using ultrasound to direct microbubbles inside the body towards targets of interest with the purpose of carrying drugs, binding them to specific tissues to aid medical imaging techniques, or to exploit locally the effects of cavitation by making bubbles collapse at precise locations.

In sonochemistry, the appearance of bubbles is usually desirable to promote mixing and to produce localized high temperatures through cavitation. These bubbles have been observed to move in complicated ways in response to sound waves \citep{Rensen_etalPRL2001}, in part due to the not-so-well understood effect of the history force \citep{ToegelLutherLohsePRL2006}.

Another modern technology where the acoustic transport of bubbles could be of interest is the production of hydrogen through artificial photosynthesis \citep{Modestino_etalEES2016}. In fact, one of the major problems of this technology is the so-called bubble control, that is, how to transport hydrogen bubbles away from the photo-active area once they reach a certain size. One interesting possibility concerns using short acoustic pulses to drive bubbles to desired harvesting locations.

As a last example, and although its implications are still a matter of controversy, we would like to mention the growth of gas bubbles in magna caused by seismic waves \citep{MangaBrodskyAREPS2006}. Indeed, we think that the topic is suggestive and though-provoking enough to deserve a comment here. It the late 1990's, \cite{Sturtevant_etalJGR1996} and \cite{Brodsky_etalJGR1998} suggested that the triggering of volcanic eruptions by distant earthquakes may be caused by rectified diffusion of gas bubbles in magna caused by traveling seismic waves. Although they took into account rectified diffusion, they did not considered the effect of gas accumulation by coalescence caused by bubble-bubble attraction or bubble-wall interaction. Properly accounting for these effects requires modeling bubble translation in response to short pressure waves.

Motivated by these applications, several authors have studied the problem of the translation of a bubble driven by pulsating pressure waves. A first approach to the problem is phenomenological: that is, to write down a set of equations for the bubble translational dynamics assuming that they are driven by a balance between different effects: drag, inertia of the surrounding liquid (added mass), and Bjerknes force; each of them modeled using an ad-hoc expression. For instance, \citet{Rensen_etalPRL2001} put together a model of this kind to explain q the spiraling behaviour observed in bubbles immersed in an expanding pipe when subject to an acoustic forcing. \citet{Dayton_etalJASA2002} used a similar strategy to describe the motion of a coated bubble in a small pipe, thus the effect of nearby walls must be retained.

A more formal approach was followed by \citet{ToilliezSzeriJASA2008} who used the method of multiple scales to separate the component of the bubble velocity pulsating at the forcing frequency from the slow drift that yields the net contribution to the mean bubble velocity. Although the authors started their derivation from the unsteady equations of motion for a compressible sphere in the limit of large Reynolds numbers \citep[see][]{MagnaudetLegendrePoF1998}, they restricted their analysis to the case where the bubble oscillates in a permanent way, once any transient behaviour has damped out.

However, in many practical situations, the duration of the acoustic forcing is not sufficiently large to achieve steady conditions, i.e. a steady mean velocity. In particular, in the case of coated bubbles used in medical applications, short insonation pulses (of the order of 10) are desirable to improve the bubble durability and long-term stability \citep{Chen_etalJASA2003}. In bubble control in artificial photosynthesis, reducing the duration of the pulse while optimizing bubble translation is essential to avoid spending too much of the generated energy in producing the vibration. In these examples, transient effects must be taken into account during the whole duration of the pulse and the subsequent deceleration stage. As we will show in this work, this duration may be as long as several tens or even hundreds of times the viscous time based on the bubble size. Consequently, among other effects, the memory integral appearing in the bubble translation equations \citep{MagnaudetLegendrePoF1998} must be accounted for.

In fact, there are several works in the literature that point out the importance of memory effects in the acoustically-driven motion of bubbles. For instance, \citet{ToegelLutherLohsePRL2006} proved that the inclusion of this term was essential to explain the non-periodic, three-dimensional orbits of a bubble around its theoretical equilibrium position observed in single-bubble sonoluminesce experiments when the viscosity of the liquid was increased. In the context of the bubble contrast agents used in medical imaging, \cite{Garbin_etalPoF2009} reported that the trajectory of a microbubble driven by secondary Bjerknes force (the one exerted between two nearby bubbles) could not be properly described unless the history term was included in the dynamic equations.

With these ideas in mind, the aim of this paper is to present a consistent treatment of the acoustically-driven bubble dynamics taking into account transient effects. More specifically, we start from the formulation developed by \cite{MagnaudetLegendrePoF1998}, based on that by \cite{MaxeyRileyPoF1983} but with the viscous terms as given by \cite{YangLealPoF1991}, which are valid for a clean gas bubble in the limit of very viscous flow. Then, in order to derive simplified expressions for the evolution of the bubble velocity, we adopt the assumptions that the amplitude of the volume oscillations is small and their period short when measured with the bubble's viscous time, namely $t_v = R_0^2/9\nu$, where $R_0$ is the initial bubble radius and $\nu$ the kinematic viscosity of the liquid.

Although, strictly speaking, the methodology followed here is only applicable in the limit of very viscous flows, we present some experimental evidences suggesting that it is more accurate to retain the memory term and assume zero Reynolds number rather than to neglect this term and model the viscous drag with a constant Stokes' friction coefficient even if it retains finite-Reynolds effects. This extends the implications of the study to the applications mentioned at the beginning of this section. In fact, although some of them (such as medical imaging) may lay outside the limits of validity of the theory, still the main conclusion holds: the memory force should not be neglected in realistic acoustic pulses.

The paper is thus structured as follows: section \ref{sec:experimental_evidences} shows some experimental evidences that point out the importance of properly accounting for memory effects in the dynamics of bubble translation. Section \ref{sec:problem_formulation} is devoted to the statement of the problem, and in particular to the presentation of the bubble dynamic equation that will be used in this work. The algorithm used to integrate this equation numerically is then presented in section \ref{sec:numerical_method}, whereas the asymptotic solution is derived in section \ref{sec:asymptotic_solution}. Some representative results, obtained through the methodology described previously, are discussed in section \ref{sec:discussion}. Finally, some conclusions are sketched in section \ref{sec:conclusions}.

\section{Experimental evidences\label{sec:experimental_evidences}}

Before starting with the formulation of the problem, it is illustrative to introduce an experimental observation that points out the need to account for the memory effects in the translation of bubbles driven by short acoustic pulses.

\begin{figure}
    \centering
    \includegraphics[width=\textwidth]{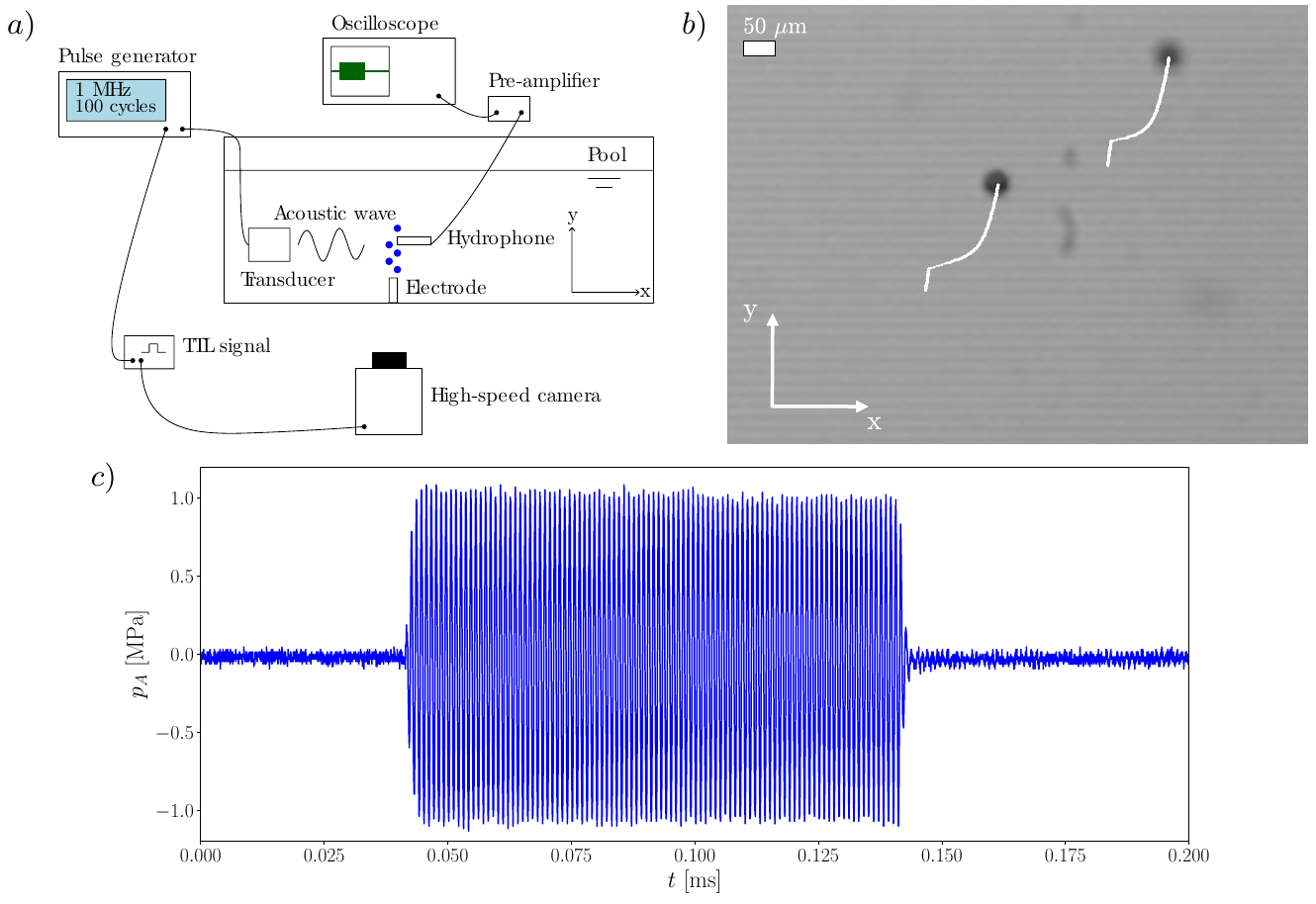}
    \caption{\label{fig:experimental_setup} (\textit{a}) Sketch of the experimental set-up. The dimensions of the pool are 100 $\times$ 40 $\times$ 40 cm$^3$ and the distance between electrode and transducer is 6.5 cm. The frame rate of the movie (acquired with a high-speed camera MEMRECAM HX-3) is 100,000 fps. (\textit{b}) Frame of a high-speed movie which illustrates the trajectory followed by two bubbles, with equilibrium radii around $R_0=20 - 25\,{\mu}\mathrm{m}$, rising in water and traveling in the horizontal direction under the effect of an ultrasound pulse. ($c$) Acoustic signal corresponding to the experiment shown in ($b$).}
\end{figure}

We generate isolated hydrogen bubbles with radii, $R_0$, in the range between 15-25 $\mu$m by water electrolysis. The cathode is placed at the bottom of a transparent acrylic tank with dimensions 100 $\times$ 40 $\times$ 40 cm$^3$. When the bubbles are far enough from the electrode, an ultrasound cylindrical 1-inch transducer with a central frequency of 1 MHz (Sonatest PIM1.0), powered by a pulser-receiver (Ritec RPR-4000), sends a single acoustic pulse of 100 cycles horizontally towards the rising bubble. The usage of a single pulse precludes the occurrence of steady streaming \citep{RileyARFM2001}. This has been checked by adding a low concentration of small tracer particles in some selected experiments. A sketch of the setup is shown in figure \ref{fig:experimental_setup}(\textit{a}), and more details about the experimental technique can be found in \cite{MedinaPalomoPhDThesis2015}.
Upon finalization of the experiment, we place the tip of a hydrophone (Ondacorp HGL-0400 with pre-amplifier AH-2010-025) in the field of view of the camera and repeat the acoustic pulse to measure the pressure wave (figure \ref{fig:experimental_setup}$c$).

\begin{figure}
\centering
\includegraphics[width=\textwidth]{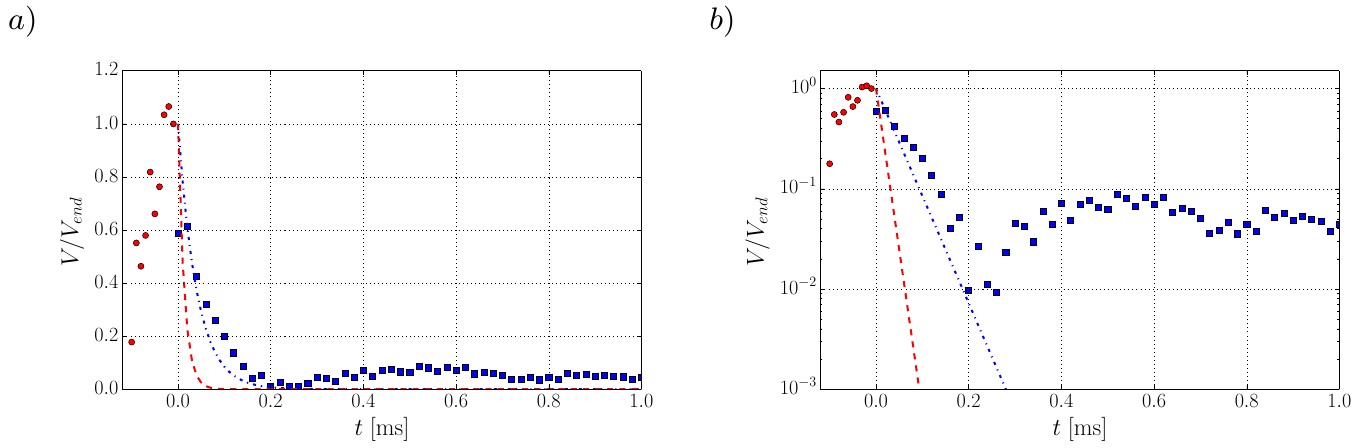}
\caption{\label{fig:experimental_velocity}(\textit{a}) Experimental velocity of a bubble, with radius $R_0 = 15.6$ ${\mu}$m (experiment 5 of table \ref{table_exp_bubbles}), along the direction of propagation of the acoustic wave. Red circles are velocities measured during the acoustic pulse and blue squares velocities during the deceleration stage. The dash-dotted lines represent the deceleration calculated assuming a constant drag coefficient, thus $F_\mathrm{drag} = \Cd \mu R_0 V$, which yields $V/V_0 = \exp(-3 \Cd \nu t / 2\pi R_0^2)$ for the evolution of the velocity. Red dashed: $\Cd = 12\pi$, Blue dash-dotted: $\Cd = 4\pi$. (\textit{b}) Same plot, but in semi-logarithmic coordinates, to evidence that no single value of $\Cd$ can describe well the experimental data, as the bubble sustains its velocity for a very large time.}
\end{figure}

In response to the acoustic excitation, the bubble departs from its otherwise vertical trajectory moving along the direction of propagation of the acoustic wave, as shown in figure \ref{fig:experimental_setup}(\textit{b}). Using digital image processing on a high-speed video of the translating bubble acquired with a Memrecam HX-3 operating at 100,000 fps and synchronized with the pulser), we track the trajectory of its center, which allows us to compute the bubble's horizontal ($x$ component) velocity. A typical plot of this velocity for a pulse is depicted in figures \ref{fig:experimental_velocity}(\textit{a}) and \ref{fig:experimental_velocity}(\textit{b}). We report in table \ref{table_exp_bubbles} the conditions of five experiments corresponding to bubbles of different sizes translating under the insonation pulse described above. The mean velocity $V_0$ reported in the table is computed using the bubble locations measured along the duration of the pulse (0.1 ms, which, recording at 100,000 fps, yields 10 frames).
\begin{table}
\centering
\begin{tabular}{ccccc}
Bubble & $R_0$ ($\mu$m) & $V_0$  (m/s) & $t_v$ ($\mu$s) & $Re = V_0 R_0 / \nu$ \\
1 &  20.6 & 0.146 & 47.2 & 3.0 \\
2 & 13.4 & 0.104 & 20.0 & 1.4 \\
3 & 25.4 & 0.187 & 71.7 & 4.8 \\
4 & 22.0 & 0.225 & 53.8 & 5.0 \\
5 & 15.6 & 0.097 & 27.2 & 1.5
\end{tabular}
\caption{\label{table_exp_bubbles}Experimental conditions of the bubble translation experiments. For all cases, the acoustic pulse is the same: $f = 1$ MHz, 100 cycles and $p_A = 1$ MPa.}
\end{table}

For reasons that will be explained in the discussion \S\ref{sec:comparison_experiments},
we focus here on the deceleration of the bubble upon the finalization of the pulse. In figure \ref{fig:experimental_velocity} we compare the bubble velocity measured in one of our experiments with the predictions obtained computing the viscous drag with Stokes' formula:
\begin{equation}
\vec{F}_\mathrm{drag} = -\Cd \mu R_0 \vec{V},
\label{eq:stokes_drag}
\end{equation}
where $\mu$ is the liquid dynamic viscosity, $R_0$ the equilibrium bubble radius and $\vec{V}$ the mean translational velocity of the bubble. In particular, the equation to be solved is:
\begin{equation}
\frac{1}{2}\left(\frac{4\pi}{3}\rho R_0^3\right) \frac{\mathrm{d}\vec{V}}{\mathrm{d}t} = - C_d\mu R_0 \vec{V}.
\end{equation}
Notice that the inertia of the bubble is not given by the (usually negligible) mass of its gas contents, but rather by the added or virtual mass, which is one half of the mass of liquid displaced by a spherical bubble \citep{LeightonBOOK}.
The comparison has been done for two values of the drag coefficient, $\Cd$, namely 4$\pi$ and 12$\pi$. The former would correspond to a clean bubble moving steadily at $\Rey = 0$, and it is the smallest value that this coefficient can have. The latter is the value calculated by \cite{MagnaudetLegendrePoF1998}, valid for short times after the deceleration starts.
In figure \ref{fig:experimental_velocity}(\textit{b}), where the $y$-axis is plotted in logarithmic scale, it is shown that the experimental values for the deceleration of the bubble cannot be fitted by a straight line. This means that the bubble velocity does not decay exponentially and therefore, the drag coefficient cannot be well described by a constant.

In the light of these results, we can state that the experimental evidence points out that the drag felt by the bubble cannot be modeled by adopting any single value of the Stokes' drag coefficient. Therefore, we must consider the complete expression of the viscous force, including the history integral. At the end of section \ref{sec:discussion} we will come back to the analysis of the experiments described here.
 
\section{Problem formulation}\label{sec:problem_formulation}

We consider here the motion of a clean gas bubble of equilibrium radius $R_0$ immersed in a flow that, far away from the bubble, is given by the velocity and pressure fields, $\vec{v}_\infty(x,t) = v_\infty(x,t)\vec{e_x}$ and $p_\infty(x, t)$ respectively. In particular, we focus on the case where this far field corresponds to an acoustic pressure wave, thus velocity and pressure are related by
\begin{equation}
\frac{\partial \vvinf}{\partial t} = -\frac{1}{\rho}\vec{\nabla}p_\infty.
\label{eq:acoustic_equation}
\end{equation}
Notice that, although strictly speaking the velocity field given by equation (\ref{eq:acoustic_equation}) is not incompressible, the fact that it changes along distances of the order of the wavelength, $\lambda$, makes it possible to replace the real compressible flow field by an unsteady but uniform outer flow around the bubble, provided its radius is much smaller than the wavelength, $R_0/\lambda \ll 1$. This assumption, which will be adopted throughout this work, is equivalent to equation (23) of \cite{MaxeyRileyPoF1983}, and effectively implies that the outer flow changes over distances much longer than the size of the particle. Additionally, as will be shown below, the smallness of the amplitude of the pressure forcing would also make possible to neglect the nonlinear terms, as they are quadratic in a small parameter.

The starting point for our derivation is the expression given by \cite{MagnaudetLegendrePoF1998} for the force exerted by the uniform flow of a liquid, with velocity $\vvinf(t)$, on a bubble with a time-varying radius $R(t)$ that moves with velocity $\vec{V}(t)$ inside the flow. In fact, for a bubble with negligible mass this force must vanish. Consequently, assuming that the Reynolds numbers based on both the bubble's relative velocity with respect to the fluid and on the wall speed are small, $R_0 \left|v_\infty - V\right|/\nu \ll 1$ and $R_0 \dot{R} / \nu \ll 1$ respectively, we have
\begin{eqnarray}
0&=& -\frac{4}{3}\pi R^3 \vec{\nabla}p_\infty + \frac{1}{2}\left[ \frac{4}{3}\pi\rho \dfi{\left[R^3\left(\vvinf-\vec{V}\right)\right]}{t} \right] +
4\pi \nu \rho R\left(\vvinf-\vec{V}\right) \nonumber\\
& &+\,8\pi \nu \rho \int\limits_{0}^{t} \exp\left[9\nu \int\limits_{\bar{t}}^{t} R(s)^{-2}\rmd s\right]\erfc\left[ \sqrt{9\nu\int\limits_{\bar{t}}^{t} R(s)^{-2}\rmd s}\, \right] \times \nonumber\\
& & \times \, \dfi{\left[ R(\bar{t}) \left( \vvinf(\bar{t})-\vec{V}(\bar{t}) \right) \right]}{\bar{t}}\,\rmd \bar{t},\label{eq:dynamic_equation_dimensions}
\end{eqnarray}
where the total viscous contribution is given by the last two terms, namely the quasi-steady drag force and the memory integral force. Hereafter, to simplify the notation, the time dependence will be omitted except inside the memory integral term. Notice also that the liquid velocity used in this equation is that particularized at the location of the bubble, so it does not depend on the location anymore. For this reason, we write $\mathrm{d}\vec{v}_\infty/\mathrm{d}t$ instead of $\partial\vec{v}_\infty/\partial t$.
 
Besides an initial condition for the velocity, to compute the translational response of the bubble, equation (\ref{eq:dynamic_equation_dimensions}) must be completed with an equation relating the bubble's volume with the pressure, such as the Keller-Miksis equation \citep{KellerMiksis1980}:
\begin{multline}
\left( 1-\frac{\dot{R}}{c} \right)R\ddot{R} + \frac{3}{2}\dot{R}^2\left( 1-\frac{\dot{R}}{3c} \right) =
 \left( 1+\frac{\dot{R}}{c} \right)\frac{1}{\rho} \left( p_\mathrm{g} -p_0 \right) + \frac{R\,\dot{p_\mathrm{g}}}{\rho c}\\
- \frac{4\mu}{\rho}\left( \frac{\dot{R}}{R} + \frac{\ddot{R}}{c} \right) - \frac{2\sigma}{\rho R}
 -\left( 1+\frac{\dot{R}}{c} \right)\frac{1}{\rho}p_\mathrm{ac}\left( \omega ( t - \vec{n}\cdot\vec{x}/c) \right)
\label{eq:KellerMiksis}
\end{multline}
where $p_\mathrm{g}$ is the gas pressure inside the bubble, given by
\begin{equation}
p_g = \left( p_0 + \frac{2\sigma}{R_0} -p_\mathrm{v} \right)\left( \frac{R}{R_0} \right)^{-3\kappa}.
\end{equation}
Here, $c$ the speed of sound in the liquid, $\sigma$ the surface tension, $\kappa$ the polytropic exponent of the gas evolution inside the bubble, $p_0$ the ambient pressure far away from the bubble in the absence of acoustic waves, and $p_\mathrm{v}$ the vapour pressure inside the bubble, which will be neglected henceforth. Finally, the acoustic pressure at the bubble surface is represented by $p_\mathrm{ac} = p_\infty - p_0$. Notice that we have already introduced here the assumption that the pressure corresponds to that of a traveling wave propagating along the direction given by the unit vector $\vec{n} = \vec{e_x}$, by letting it depend on $\xi = \omega\left(t - \vec{n}\cdot\vec{x}/c\right)$. It should be pointed out that, in virtue of the hypothesis $R_0/\lambda \ll 1$, a phase-lag $\omega R_0/c$ is neglected in $\xi$. Nonetheless, in the rest of Equation (\ref{eq:KellerMiksis}), terms of order $R_0/\lambda$ are retained, as they may have a significant contribution in damping bubble volume oscillations \citep{ProsperettiJASA1977}.

%
%
\subsection{Dimensionless formulation}

Keeping in mind the motivation of our work, namely the effect of the memory integral term on the translation of bubbles in response to a short acoustic wave, we use the viscous time of the problem, $R_0^2/9\nu$, to define a dimensionless time variable $\tau = 9\nu t/R_0^2$. It will become clear later that the choice of this time scale will simplify the evaluation of the memory integral. The equilibrium radius of the bubble, $R_0$, is used as length scale, so $\tilde{\vec{x}} = \vec{x}/R_0$ and $a = R/R_0$. Consistently with this formulation, the pressure field can be expressed as
\begin{equation}
p_{\rm ac} = p_A\,f(\Sto \, \tau - M \vec{n}\cdot \vec{\tilde{x}}),
\label{eq:dimensionless_acoustic_pulse}
\end{equation}
with $p_A$ being the amplitude of the pressure wave, $\Sto=\omega R_0^2/9\nu$ the Stokes number and $M = \omega R_0/c$ the Mach number. Notice that, under the assumptions adopted here, $M \ll 1$, since $M \sim R_0 / \lambda$. It is also worth mentioning that the Stokes number that emerges in the problem can be interpreted as the ratio between the viscous time and the period of the acoustic wave.

Introducing equation (\ref{eq:dimensionless_acoustic_pulse}) into (\ref{eq:acoustic_equation}) and making use of the characteristic scales defined above, the corresponding dimensionless equation for the external flow reads
\begin{equation}
\dfr{\vvinft}{\tau} = \varepsilon M \Sto \, \vec{n} \, \frac{\rmd}{\rmd\tau}f(\Sto \, \tau - M \vec{n}\cdot \vec{\tilde{x}}),
\label{eq:dimensionless_acoustic_equation}
\end{equation}
where the parameter $\varepsilon = p_A/\rho\,\omega^2 R_0^2$ modulates the amplitude of the bubble oscillations.
Notice that the spatial derivative found in (\ref{eq:acoustic_equation}) has been transformed into a temporal one, which will come in handy below in order to integrate the equation.

Similarly, the dimensionless momentum equation yields
\begin{equation}
0 = \varepsilon M \Sto a^3\vec{n} \dfr{f(\xi)}{\tau} + \frac{1}{2}\dfr{(a^3 \vUt)}{\tau} + \frac{1}{3} a \vUt + \frac{2}{3} \int_0^\tau G(\tau, \bar{\tau}) \dfr{( a \vUt)}{\bar{\tau}} \rmd\bar{\tau},
\label{eq:dimensionless_momentum_eq}
\end{equation}
where we have defined the history kernel as:
\begin{equation}
G(\tau, \bar{\tau}) = \exp\left[\,\int\limits_{\bar{\tau}}^{\tau} a(s)^{-2}\rmd s\right]\erfc\left[ \sqrt{ \int\limits_{\bar{\tau}}^{\tau} a(s)^{-2}\rmd s}\, \right],
\label{eq:kernel_tau}
\end{equation}
and the relative velocity $\vvinft-\tilde{\vec{V}}$ is hereafter denoted by $\vUt$.

Additionally, to complete the problem, the dimensionless Keller-Miksis equation is required
\begin{eqnarray}
& & \ddot{a} \left( \frac{4}{9}\frac{M}{\Sto}-\frac{M}{\Sto}a\dot{a}+a \right) =  \frac{1}{2}\frac{M}{\Sto}\dot{a}^3 + M \Sto\dot{a}\left[ \left( \Pi_0 +\frac{1}{\We} \right) a^{-3\kappa} - \Pi_0 \right]-\frac{3}{2}\dot{a}^2 \nonumber \\ 
& &-\frac{4}{9} \frac{\dot{a}}{a}- \frac{\Sto^2}{\We}\frac{1}{a} 
   + \left( \Pi_0+\frac{1}{\We} \right)\Sto^2 a^{-3\kappa} - \Pi_0 \Sto^2
 -3M\Sto\kappa\dot{a}\left( \Pi_0 + \frac{1}{\We} \right)a^{-3\kappa} \nonumber \\
& & -\left( 1+ \frac{M}{\Sto}\dot{a} \right)\Sto^2\varepsilon f\left(\xi\right)
\label{eq:dimensionless_keller_miksis}
\end{eqnarray}
with $\Pi_0=p_0/\rho R_0^2\omega ^2$ being the dimensionless ambient pressure (Euler number) and $\We=\rho R_0^3\omega^2/2\sigma$ the Weber number.

As a last step, we rescale the time in order to simplify the treatment of the history integral \citep{MagnaudetLegendrePoF1998}. This simplification will come in particularly handy for the implementation of the numerical method. Indeed, defining a rescaled time such that, $\rmd\tau^* = \rmd\tau/a^2(\tau)$, the kernel (\ref{eq:kernel_tau}) results
\begin{equation}
G(\tau, \bar{\tau}) = G^*(\tau^*-\bar{\tau}^*) = \exp\left[ \tau^*-\bar{\tau}^* \right]\erfc\left[ \sqrt{ \tau^*-\bar{\tau}^*} \right].
\label{eq:kernel_tau_star}
\end{equation}
Hereafter, this new time variable will be referred to as non-linear time. After some algebra, equations (\ref{eq:dimensionless_acoustic_equation}), (\ref{eq:dimensionless_momentum_eq}) and (\ref{eq:dimensionless_keller_miksis}) read
\begin{equation}
\dfr{\vvinft}{\tau^*} = \varepsilon M \Sto \vec{n} \dfi{f(\Sto\,\tau - M\vec{n}\cdot \vec{\tilde{x}})}{\tau^*}
\label{eq:acoustic_equation_rescaled_time}
\end{equation}
\begin{equation}
0 = \varepsilon M \, \Sto \, a\, \vec{n} \dfr{f(\xi)}{\tau^*} + \dfr{(\ln a)}{\tau^*}a\vUt 
+ \frac{1}{2}\dfr{(a\vUt)}{\tau^*} +\frac{1}{3} a\vUt + 
\frac{2}{3} \int\limits_{0}^{\tau^*} G^*(\tau^*-\bar{\tau}^*)\dfr{(a\vUt)}{\bar{\tau}^*}\rmd\bar{\tau}^*
 \label{eq:dynamic_equation_rescaled_time}
\end{equation}
\begin{eqnarray}
& & \ddot{a}^* \left( \frac{4}{9}\frac{M}{\Sto}-\frac{M}{\Sto}\frac{\dot{a}^*}{a}  +a \right) =  
2\frac{(\dot{a}^*)^2}{a}\left( \frac{4}{9}\frac{M}{\Sto} -\frac{M}{\Sto}\frac{\dot{a}^*}{a} + a \right)
+ \frac{1}{2} \frac{M}{\Sto} \frac{(\dot{a}^*)^3}{a^2} \nonumber\\
& &  +M \Sto\dot{a}^*a^2\left[ \left( \Pi_0 + \frac{1}{\We}\right)a^{-3\kappa} - \Pi_0 \right]
-\frac{3}{2}\left( \dot{a}^* \right)^2  -\frac{4}{9}a\dot{a}^*-\frac{\Sto^2}{\We}a^3 \nonumber\\
& & -\Pi_0 \Sto^2 a^4 + \left( \Pi_0 +\frac{1}{\We}\right)\Sto^2 a^{-3\kappa+4}
 -3M\Sto\kappa a^2 \dot{a}^* \left( \Pi_0 + \frac{1}{\We} \right) a^{-3\kappa} \nonumber\\
& & -\left( 1+\frac{M}{\Sto}\frac{\dot{a}^*}{a^2} \right)\Sto^2 a^4 \varepsilon f(\xi),
\label{eq:keller_miksis_equation_rescaled_time}
\end{eqnarray}
where $\dot{a}^* $ and $\ddot{a}^*$ denote the first and the second derivatives of $a$ with respect to the rescaled time $\tau^*$.

%
%
\subsection{Some remarks on the underlying hypotheses\label{sec:hypotheses}}

We conclude this section with a few comments about the hypotheses adopted thus far and some of their implications. Firstly, as is customarily done in acoustics, the Mach number is assumed to be very small, $M \ll 1$, which here implies that the size of the particle is much smaller than the length scale of the outer flow, as $R_0/\lambda = M/(2\pi)$. Following \cite{MaxeyRileyPoF1983}, this allows us to neglect the Faxen correction terms in Equation (\ref{eq:dynamic_equation_dimensions}) and to approximate 
$\mathrm{D}\vvinf/\mathrm{D}t \approx \partial\vvinf/\partial t$ in (\ref{eq:acoustic_equation}).

Secondly, we require the Reynolds numbers based on both the relative velocity between the bubble and the liquid, and on the bubble's wall velocity, to be small. The smallness of the former justifies the expressions used for the viscous drag (both the steady and history-dependent components), a condition also employed by \cite{MaxeyRileyPoF1983}. To express these conditions in terms of the dimensionless parameters defined here, $\varepsilon$, $M$, and $\Sto$, we can use order of magnitude analysis on equation (\ref{eq:acoustic_equation}) to get $O(\vvinf) \sim \varepsilon M \omega R_0$. Assuming that the bubble's relative velocity, $\vvinf-\vec{V}$, is as much of this order, the condition of small translational Reynolds number reads
\begin{equation}
\frac{R_0\left|\vvinf-\vec{V}\right|}{\nu} \sim \varepsilon\,M\,\Sto \ll 1.
\label{eq:small_translational_Reynolds}
\end{equation}
We consider now the condition of small wall-based Reynolds number. The order of magnitude of the amplitude of the radial oscillations is going to denoted by $\delta R_0$. When the insonation frequency is far from resonance, $\delta \sim \varepsilon$, whereas when the bubble is excited close to its natural frequency we will have $\delta \sim \varepsilon \Sto$ (see Equation (\ref{eq:simplified_radius}) below). Using this parameter, the condition of small wall-based Reynolds number may be written:
\begin{equation}
\frac{R_0 \dot{R}}{\nu} \sim \delta\,\Sto \ll 1,
\label{eq:small_wall_Reynolds}
\end{equation}
which translates into $\varepsilon\Sto \ll 1$ in the former case and $\varepsilon\Sto^2 \ll 1$ in the latter. Notice that, close to resonance, this condition results very restrictive at first sight, as it strongly limits the amplitude of pressure wave. Nonetheless, numerical simulations that will be reported elsewhere \citep{RodriguezRodriguez_preparation} suggest that the expressions used here can be applied at least for wall-based Reynolds numbers of order unity. The reason is that the motion induced by the pulsation of the wall is radial and therefore does not generate vorticity. Consequently, although this motion is responsible for part of the virtual mass, its influence on the viscous drag is very limited, at least for order-unity wall-based Reynolds numbers.

Finally, a third restriction imposed by \cite{MaxeyRileyPoF1983} in order to neglect advection effects in the flow around the particle is that $R_0^2 \left|\vvinf \right|/(\nu\lambda) \ll 1$, which, using the notation defined here, turns into
\begin{equation}
\frac{R_0^2\left|v_\infty\right|}{\nu\lambda} \sim \varepsilon\,M^2 \Sto \ll 1.
\end{equation}
Notice that, effectively, this condition does not imply any further restriction beyond (\ref{eq:small_wall_Reynolds}) and $M \ll 1$. 

Furthermore, although not needed to formulate the problem, the asymptotic approach derived in section \ref{sec:asymptotic_solution} will additionally require the bubble radial oscillations to be of small amplitude, $\delta \ll 1$, and that their frequency is fast compared to the inverse of the viscous time. In other words, that the Stokes number is large, $\Sto \gg 1$.

\section{Numerical method}\label{sec:numerical_method}

The translational response of the bubble due to the acoustic pulse is given by equations (\ref{eq:acoustic_equation_rescaled_time}), (\ref{eq:dynamic_equation_rescaled_time}) and (\ref{eq:keller_miksis_equation_rescaled_time}). They constitute a system of coupled, non-linear, ordinary integro-differential equations that must be solved numerically.

The history integral is computed with the numerical method proposed by \cite{KimSirignano98}, which provides an accurate approximation even when the kernel becomes singular at one of the limits (as is the case for the solid-sphere kernel). More specifically, at the $n$-th time step, the integral is approximated by:
\begin{eqnarray}
& &\int\limits_{0}^{\tau^*}G^*(\tau^*-\bar{\tau}^*)\vec{\tilde{W}}_{\bar{\tau}^*}\,\rmd\bar{\tau}^* = 
\int\limits_{0}^{n\Delta\tau^*}G^*(\tau^*-\bar{\tau}^*)\vec{\tilde{W}}_{\bar{\tau}^*}\,\rmd\bar{\tau}^* \approx \nonumber\\
\approx \frac{\Delta\tau^*}{6}\sum\limits_{i=1}^{n-1}\biggl[ & &G^*(n\Delta\tau^*-(i-1)\Delta\tau^*)\left.\vec{\tilde{W}}_{\tau^*}\right| _{i-1}  + \nonumber\\
& & + 2G^*(n\Delta\tau^*-(i-0.5)\Delta\tau^*)\left(\left.\vec{\tilde{W}}_{\tau^*}\right|_{i-1}+\left.\vec{\tilde{W}}_{\tau^*}\right|_{i}\right) + \nonumber\\
& & + G^*(n\Delta\tau^*-i\Delta\tau^*)\left.\vec{\tilde{W}}_{\tau^*}\right|_{i} \biggr] + \nonumber\\
& & + \frac{0.1\Delta\tau^*}{2}\biggl[ \frac{8\sqrt{2}}{3}G^*(0.05\Delta\tau^*)\left.\vec{\tilde{W}}_{\tau^*}\right|_{n}-
\frac{4}{3}G^*(0.1\Delta\tau^*)\left.\vec{\tilde{W}}_{\tau^*}\right|_{n} \biggr] + \nonumber\\
& & +\frac{0.9\Delta\tau^*}{6}\biggl[ G^*(\Delta\tau^*)\left.\vec{\tilde{W}}_{\tau^*}\right|_{n-1} + 
2G^*(0.55\Delta\tau^*)\left(\left.\vec{\tilde{W}}_{\tau^*}\right|_{n-1}+\left.\vec{\tilde{W}}_{\tau^*}\right|_{n}\right) + \nonumber\\
& & + G^*(0.1\Delta\tau^*)\left.\vec{\tilde{W}}_{\tau^*}\right|_{n} \biggr].
\label{eq:Numerical_method_history_term}
\end{eqnarray}
Here, $\rmd(a\vUt)/\rmd\tau^*$ has been denoted by $\vec{\tilde{W}}_{\tau^*}$ for simplicity. Notice that this method requires the usage of a constant time step, $\Delta\tau^*$. 

Consistently with this, the time marching has been performed with a fourth-order Adams-Bashforth algorithm, which is a multi-step explicit method. The multi-step character of the method facilitates the computation of the memory term, since the algorithm used for its evaluation (Equation \ref{eq:Numerical_method_history_term}) uses the current, $n$, and previous values. More specifically, the algorithm yields:
\begin{eqnarray}
\vec{y}_{n+1} & = & \vec{y}_n + \Delta\tau^*\,\dot{\vec{y}}^*(\tau^*_n, \vec{y}_n)\\
\vec{y}_{n+2} & = & \vec{y}_{n+1} + \Delta\tau^*\left( \frac{3}{2}\dot{\vec{y}}^*(\tau^*_{n+1}, \vec{y}_{n+1})-\frac{1}{2}\dot{\vec{y}}^*(\tau^*_{n}, \vec{y}_{n}) \right)\\
\vec{y}_{n+3} & = & \vec{y}_{n+2} +  \Delta\tau^*\left( \frac{23}{12}\dot{\vec{y}}^*(\tau^*_{n+2}, \vec{y}_{n+2}) - \frac{4}{3}\dot{\vec{y}}^*(\tau^*_{n+1}, \vec{y}_{n+1})+\frac{5}{12}\dot{\vec{y}}^*(\tau^*_{n}, \vec{y}_{n}) \right)\\
\vec{y}_{n+4} & = & \vec{y}_{n+3} +  \Delta\tau^*\left( \frac{55}{24}\dot{\vec{y}}^*(\tau^*_{n+3}, \vec{y}_{n+3}) - \frac{59}{24}\dot{\vec{y}}^*(\tau^*_{n+2}, \vec{y}_{n+2})\right. \nonumber \\
 && \hspace{134.7pt} \left. +\, \frac{37}{24}\dot{\vec{y}}^*(\tau^*_{n+1}, \vec{y}_{n+1})-\frac{3}{8}\dot{\vec{y}}^*(\tau^*_{n}, \vec{y}_{n}) \right).
\label{eq:Adams_Bashforth_method}
\end{eqnarray}
Here, $\vec{y} = \left\{a, \dot{a}^*, \vec{\tilde{W}}\right\}$ is the vector of unknowns. In the simulations reported in this work, the time step chosen is in the range of $\Delta\tau^* = 2\cdot 10^{-4} - 9\cdot 10^{-4}$. In all of the simulations presented here, we have tested for the temporal convergence of the numerical integration by reducing the time step up to $500$ times.

\section{Asymptotic solutions to the bubble translation problem\label{sec:asymptotic_solution}}

As discussed in the previous section, including the history force in the computation of the bubble velocity leads to an integro-differential equation with an integral term that must be evaluated numerically at each time step. Thus, the calculation of the bubble response under a typical acoustic pulse, which might require to integrate for a large number of acoustic cycles, results computationally expensive. With this motivation, we develop in this section an analytical solution for the time evolution of the bubble velocity. Even more importantly, the analytical solution will allow us to obtain physical insight on the physics of the problem and to elucidate which terms are dominant in the bubble dynamics.

We start by assuming that the bubble undergoes small-amplitude volume oscillations, which requires $\delta \ll 1$. As mentioned in subsection \ref{sec:hypotheses}, this translates into $\varepsilon\Sto \ll 1$ when the insonation takes place close to the resonance frequency and into $\varepsilon \ll 1$ otherwise. This poses a difficulty, as the small parameter used in the asymptotic expansions depends on other parameters of the problem (the Stokes number and the dimensionless resonance frequency, $\Omega_0$, defined below). To avoid this difficulty, and to derive a single formulation that can be used uniformly regardless of the insonation frequency, we expand the bubble radius as
\begin{equation}
a=1+\varepsilon r,
\label{eq:expansion_a}
\end{equation}
although it means that the dimensionless radius, $r$, could be as large as $\Sto$ close to resonance. Notice that this choice does not reduce the validity of the asymptotic analyses performed below, as long as the condition $\delta \ll 1$ is fulfilled.

Introducing this expression into the dimensionless Keller-Miksis equation (\ref{eq:dimensionless_keller_miksis}), and neglecting second-order and higher terms, yields:
\begin{equation}
\left(1+\frac{4M}{9\Sto}\right)\frac{\rmd^2r}{\rmd\tau^2} + \left(\frac{4}{9} + M\Sto\,\Omega_0^2\right)\frac{\rmd r}{\rmd\tau} + \Sto^2\Omega_0^2\, r = -\Sto^2 f(\Sto\tau - M\vec{n}\cdot\vec{\tilde{x}}),
\label{eq:linear_keller_miksis}
\end{equation}
with $\Omega_0^2 = 3\kappa\Pi_0 + \left(3\kappa-1\right)\We^{-1}$. Furthermore, we can expand the relative velocity times the radius, $\vWt = a\vUt=a\left(\vvinft-\tilde{\vec{V}}\right)$, assuming that its order of magnitude is dictated by that of $\vvinft$, which can be obtained by integrating equation (\ref{eq:dimensionless_acoustic_equation}):
\begin{equation}
\vvinft = \varepsilon M \Sto \, \vec{n} f(\Sto \,\tau - M\vec{n}\cdot\vec{\tilde{x}}).
\label{eq:dimensionless_acoustic_velocity}
\end{equation}
Thus, we assume that $\vWt \sim \varepsilon M St$, which suggests the following ansatz:
\begin{equation}
\vWt = \varepsilon M \Sto \left( \vWt^{(1)} + \varepsilon \vWt^{(2)} \right) + O\left( \varepsilon^3\right).
\label{eq:expansion_U}
\end{equation}
Henceforth, and as has been done when integrating (\ref{eq:dimensionless_acoustic_equation}), we will neglect the influence of the position on the phase of the acoustic wave seen by the bubble, namely the contribution $-M\vec{n}\cdot\vec{\tilde{x}}$. It can be checked that this phase shift is negligible as long as the bubble translates a distance much smaller than $\Sto/M$ over one acoustic cycle. Furthermore, we will also drop the vector character of the velocities, as they all are aligned with the direction of propagation of the acoustic pulse, $\vec{n}$.

The smallness of the amplitude of the oscillations also suggests that the difference between the nonlinear and the linear times is going to be small. In order to take advantage of this, we will rewrite Equation (\ref{eq:dimensionless_momentum_eq}) in terms of $\Wt$ and divide the resulting equation by $a^2$,
\begin{equation}
0 = \varepsilon\,M\,\Sto\,a\frac{\rmd f}{\rmd\tau} + \frac{\rmd\ln a}{\rmd\tau}\Wt + \frac{1}{2}\frac{\rmd\Wt}{\rmd\tau} + \frac{1}{3}a^{-2}\Wt + \frac{2}{3} a^{-2} \int\limits_{0}^{\tau} G^*(\tau^*-\bar{\tau}^*)\dfr{\Wt}{\bar{\tau}}\rmd\bar{\tau}
\label{eq:dynamic_equation_before_asympt}
\end{equation}
Introducing the expansions for $a$ and $\Wt$, and collecting terms of the same order, we get for the leading order, $O(\varepsilon M St)$:
\begin{equation}
0 = \frac{\rmd f}{\rmd \tau} + L\left[\Wt^{(1)}\right],
\label{eq:leading_order_eq}
\end{equation}
whereas for the first-order correction, $O(\varepsilon^2 M St)$:
\begin{equation}
0 = r \frac{\rmd f}{\rmd \tau} + \left(\frac{\rmd r}{\rmd\tau} - \frac{2}{3} r\right) \Wt^{(1)}+ L\left[\Wt^{(2)}\right].
\label{eq:first_order_eq}
\end{equation}
In these equations we have made use of the following linear, but non-local in time, operator:
\begin{equation}
L\left[ \cdot \right] = \frac{1}{2}\dfr{\left[ \cdot \right]}{\tau} + \frac{1}{3} \left[ \cdot \right] + \frac{2}{3}\int\limits_{0}^{\tau} G(\tau-\bar{\tau}) \dfr{\left[ \cdot \right]}{\bar{\tau}}\rmd\bar{\tau}.
\label{eq:linear_operator}
\end{equation}
It should be pointed out that a term proportional to $r$ times the history integral evaluated for $\vWt^{(1)}$ has been neglected in (\ref{eq:first_order_eq}), as its contribution is expected to be small, as will be verified later.

In the following subsections we will derive asymptotic solutions for Equations (\ref{eq:linear_keller_miksis}), (\ref{eq:leading_order_eq}) and (\ref{eq:first_order_eq}) for the case of a purely sinusoidal acoustic forcing. In subsection \ref{sec:analytical_approximate} we present an approximate solution for an infinitely long acoustic pulse. Besides being directly applicable to describe the acceleration of a bubble in practical situations, this simplified solution will serve us to gain physical insight about which terms of the dynamic equation (\ref{eq:dynamic_equation_before_asympt}) are dominant and thus must be retained to describe the transient motion of the bubble. Then, in subsection \ref{sec:leading_order} we lay down the procedure to obtain a solution for the case of a finite-duration pulse of the form
\begin{equation}
f(\tau) = \sin\left(\Sto\, \tau \right) \left[ H(\tau) - H(\tau -\tau _p)\right].
\label{eq:Pulse_function}
\end{equation}
Here, $H(\tau)$ is the Heaviside function and $\tau_p$ the duration of the pulse, which in this study has been chosen to be an integer number of times the period of the acoustic period.
Moreover, to further simplify the solution in a limit of relevance for practical applications, the Stokes number is assumed to be large, $\Sto \gg 1$. That is, the viscous time is much larger than the period of the acoustic wave.

A comment is now in place about the form of the acoustic pulse used in this study (equation (\ref{eq:Pulse_function})). Although its amplitude is constant during the duration of the pulse, $0 \le \tau \le \tau_p$, for the sake of algebraic simplicity, the theory developed here can be modified to take into account tapering. Indeed, the solution procedure sketched in subsection \ref{sec:leading_order} allows to deal with any forcing function that can be decomposed into piecewise linear combinations of trigonometric functions. Thus, it could be straightforwardly applied to pulses tapered at the start with functions of the kind $\left(1-\cos(\pi\tau/T_t)\right)/2$ for $\tau \le T_t$, with $T_t$ the duration of the tapering, and an analogous expression for the pulse end. We must point out that although the procedure is easy from the conceptual point of view, as it only involves including more segments of the same form (sinusoidal functions) to the definition of the forcing function, the computations would become very tedious.

\subsection{\label{sec:analytical_approximate}Simplified solution for the mean velocity in response to a permanent acoustic forcing}

As a first assumption, we will consider that the bubble is immediately set to oscillate in a purely periodic fashion when the acoustic pulse starts. This is a reasonable simplification, as the transient component in the solution of Equation (\ref{eq:linear_keller_miksis}) decays exponentially whereas, as we will show below, the transient terms in the bubble acceleration decay algebraically. Thus, we will approximate the amplitude of the radial oscillations by the permanent solution to the linearized Keller-Miksis equation:
\begin{equation}
r = \frac{\varepsilon\Sto}{\sqrt{\left(\displaystyle 1+\frac{4 M}{9 \Sto} - \Omega_0^2\right)^2\Sto^2 + \beta_r^2}}\cos\left(\Sto\,\tau + \phi\right),
\label{eq:simplified_radius}
\end{equation}
where $\beta_r = 4/9 + M\Sto\Omega_0^2$ and $\tan\phi = -\left(1 + 4M/9\Sto - \Omega_0^2\right)\Sto/\beta_r$.

Similarly, to obtain $\Wt^{(1)}$ --which is needed to evaluate the second forcing term in Equation (\ref{eq:first_order_eq})-- we will also neglect the terms arising from the viscous drag in the solution of Equation (\ref{eq:leading_order_eq}), including the history integral. This is justified since they are $\Sto$ times smaller than the time derivatives that appear in the equation. Doing this, we get for the leading-order relative velocity:
\begin{equation}
\Wt^{(1)} = -2\sin\left(\Sto\,\tau\right).
\end{equation}
Although we do not show it here for the sake of conciseness, it can be checked that this expression agrees very well with the full solution after a few oscillation cycles.

Introducing these expressions into Equation (\ref{eq:first_order_eq}), and after some algebra, the forcing term may be written
\begin{equation}
r\frac{\rmd f}{\rmd\tau} + \left(\frac{\rmd r}{\rmd\tau} - \frac{2}{3}r\right)\Wt^{(1)} = \frac{\Sto^2\left(F + F_c\cos 2\Sto\,\tau + F_s\sin 2\Sto\,\tau\right)}{\sqrt{\left(\displaystyle 1+\frac{4 M}{9 \Sto} - \Omega_0^2\right)^2\Sto^2 + \beta_r^2}},
\end{equation}
with
\begin{eqnarray}
F & = & \frac{3}{2}\cos\phi,\\
F_c & = & -\frac{1}{2}\cos\phi,\\
F_s & = & -\frac{1}{2}\sin\phi.
\end{eqnarray}
These expressions reveal that the forcing can be decomposed into a constant part plus an oscillation of frequency $2\Sto$. As a last simplification, and keeping in mind that we are seeking for the non-oscillatory component of the bubble velocity, we will only consider the constant part of the forcing. Applying the Laplace transform to Equation (\ref{eq:first_order_eq}), and denoting the amplitude of the constant forcing by
\begin{equation}
A_F = \frac{\Sto^2F}{\sqrt{\left(\displaystyle 1+\frac{4 M}{9 \Sto} - \Omega_0^2\right)^2\Sto^2 + \beta_r^2}},
\end{equation}
we may write:
\begin{equation}
0 = \frac{A_F}{s} + \left(\frac{1}{2}s + \frac{1}{3} + \frac{2}{3}\frac{s}{\sqrt{s}\left(1+\sqrt{s}\right)}\right){{\tilde{\cal W}}^{(2)}},
\label{eq:laplace_transform_simplified_soln}
\end{equation}
where ${\tilde{\cal W}}^{(2)}$ is the Laplace transform of $\Wt^{(2)}$. Equation (\ref{eq:laplace_transform_simplified_soln}) can be inverted exactly, as pointed out by \cite{SYeoh2009} for an analogous case, yielding for the relative velocity:
\begin{align}
\Wt^{(2)} = A_F\left\{F_{s1} + \rme^{-\alpha\tau}\left( D_{s1}\cos(\gamma\tau)+
\frac{E_{s1}-D_{s1}\alpha}{\gamma}\sin(\gamma\tau) \right) 
 + C_{s1}\rme^{s_1\tau}\erfc\sqrt{s_1 \tau} \nonumber \right.\\
 \left. + \frac{8}{3}\left[ \left( E_{s2} - 2\alpha D_{s2} \right)\Icos(\tau, \alpha, \gamma) +
 \frac{-E_{s2}\alpha + D_{s2}(\alpha^2-\gamma^2)}{\gamma}\Isin(\tau, \alpha, \gamma)  \right]   \right\}
 \label{eq:simplified_relative_velocity}
\end{align}
Here, $C_{s1}, D_{s1,2}, E_{s1,2}, F_{s1}$, $s_1$, $\alpha$ and $\gamma$ are numerical constants that have been defined in Appendix \ref{apSimplifiedSolution}, whereas functions $\Icos(\tau, \alpha, \gamma)$ and $\Isin(\tau, \alpha, \gamma)$ are specified in Appendix \ref{apFunctions}.

Finally, we recall the definition of the radius-weighted relative velocity, $\Wt = a\left(\tilde{v}_\infty-\tilde{V}\right)$, to obtain the bubble mean velocity. Indeed, keeping in mind that $\vvinft$ has zero mean, and assuming $a \approx 1$, we get for the non-oscillatory component of the velocity:
\begin{equation}
\tilde{V}_{no} \approx -\varepsilon^2 M \Sto \, \Wt^{(2)}.
\label{eq:simplified_velocity}
\end{equation}
Notice that approximating the radius by unity does not modify the non-oscillatory part of the velocity at leading order, as $r$ is a purely oscillatory function with the approximations made in this subsection. Expressions (\ref{eq:simplified_relative_velocity}) and (\ref{eq:simplified_velocity}) are compared with the numerical solution of the full problem in figures \ref{fig:case1}, \ref{fig:case2} and \ref{fig:case3}. Also, the time evolution of the bubble radius and its time derivative is plotted in figure \ref{fig:radii}. The discussion of the behavior of this solution is postponed to \S\ref{sec:discussion}, to continue now with the derivation of a more complete asymptotic solution of the problem.

\subsection{Response to a pulse of finite duration\label{sec:leading_order}}

Here we consider the response to an acoustic pulse consisting in a sinusoidal forcing of frequency $St$ starting at $\tau = 0$ and ending at a finite time $\tau_p$, which corresponds to an integer number of oscillation periods, as defined in Equation (\ref{eq:Pulse_function}). For the sake of conciseness, and due to the length of the full solution, we have relegated its expression to Appendixes \ref{apW1}, \ref{apW2_1} and \ref{apW2_2}. In this subsection we just point out the method followed in its derivation.

We start with the solution to the leading order component, namely $\Wt^{(1)}$, which is mainly oscillatory. Applying Laplace transform to Equation (\ref{eq:leading_order_eq}), yields
\begin{equation}
0= {\cal F}(s) +\frac{1}{2} \left( s \tilde{{\cal W}}^{(1)}(s)- \Wt^{(1)}|_0 \right) + \frac{1}{3}\,\tilde{{\cal W}}^{(1)}(s) 
+ \frac{2}{3} \left( \frac{s\,\tilde{{\cal W}}^{(1)}(s)}{\sqrt{s}\left( 1+\sqrt{s}\right)} \right),
\end{equation}
with ${\cal F}(s)$ the Laplace transform of $\rmd f/\rmd\tau$, $\tilde{{\cal W}}^{(1)}$ the Laplace transform of the leading order of the radius-weighted relative velocity, $\Wt^{(1)}$, and $\Wt^{(1)}|_0$ its initial value. Isolating the Laplace transform of the radius-weighted relative velocity:
\begin{equation}
\tilde{{\cal W}}^{(1)}(s) =  \frac{ \Wt^{(1)}|_0 \left( s^2+s-\frac{2}{3} + \frac{4}{3}\sqrt{s} \right)}{\left( s-s_1 \right)\left[ (s+\alpha)^2 + \gamma^2 \right]} +
{\cal F}(s)\frac{-2 \left( s^2+s-\frac{2}{3} + \frac{4}{3}\sqrt{s} \right)}{\left( s-s_1 \right)\left[ (s+\alpha)^2 + \gamma^2 \right]}.
\label{eq:leading_order_eq_laplace}
\end{equation}
The solution has been split into two components. The former, $\Wt^{(1)}_0$, is due to the initial relative velocity, non-zero in case of having a bubble immersed in a flow that is not at rest before the acoustic forcing begins, whereas the latter, $\Wt^{(1)}_f$, is due to the acoustic forcing and thus depends on the function describing the shape of the acoustic pulse.

To keep advancing, it is necessary to specify the forcing function $f(\tau)$. Recalling Equation (\ref{eq:Pulse_function}), its derivative reads
\begin{equation}
\frac{\rmd f(\tau)}{\rmd\tau} = \Sto\cos\left( \Sto \, \tau \right) \left[ H(\tau) - H(\tau -\tau _p)\right] + \sin\left(\Sto \, \tau \right) \left[ \delta(\tau) - \delta(\tau -\tau _p)\right].
\end{equation}
Since the duration of the pulse, $\tau_p$, is chosen here to be an integer number of times the period of the acoustic wave, the term containing Dirac's deltas is zero. Therefore, $\rmd f/\rmd\tau$ will be given by  
\begin{equation}
\frac{\rmd f(\tau)}{\rmd\tau} = \Sto\cos\left(\Sto \, \tau \right) \left[ H(\tau) - H(\tau -\tau _p)\right]
\label{eq:Pulse_function_derivative}
\end{equation}
and thus
\begin{equation}
{\cal F}(s) = \frac{\Sto\,s}{s^2 + \Sto^2}\left(1-e^{-s\,\tau_p}\right).
\end{equation}
Substituting this expression into (\ref{eq:leading_order_eq_laplace}), and inverting the Laplace transforms we obtain $\Wt^{(1)}(\tau)$, whose expression is provided in Appendix \ref{apW1}.

We focus now on the non-oscillatory part of the relative velocity, which mainly arises from the first-order correction, $\Wt^{(2)}$, given by Equation (\ref{eq:first_order_eq}). Indeed, this component is the responsible for the appearance of a net force, namely the Bjerknes force.
Notice that this equation has two forcing terms: one due to the acoustic forcing and another due to the contribution of the leading-order component, $\vWt^{(1)}$. As the equation is linear in $\Wt^{(2)}$, we can search for the particular solutions associated to both contributions separately:
\begin{equation}
0 = r \, \frac{\rmd f}{\rmd \tau} + L\left[\Wt^{(2)}_1\right]
\label{eq:first_order_eq_simple_1}
\end{equation}
\begin{equation}
0 = \left(\frac{\rmd r}{\rmd\tau} - \frac{2}{3}r \right)\Wt^{(1)} + L\left[\Wt^{(2)}_2\right]
\label{eq:first_order_eq_simple_2}
\end{equation}
with $\Wt^{(2)}=\Wt^{(2)}_1 + \Wt^{(2)}_2$.

In the case of equation (\ref{eq:first_order_eq_simple_1}), the forcing term is a sum of products of decaying exponentials and trigonometric functions, modulated in time by the Heaviside step function.
Therefore, $\Wt^{(2)}_1$ can be obtained by applying the Laplace transform method as explained in subsection \ref{sec:analytical_approximate}.

However, in the case of equation (\ref{eq:first_order_eq_simple_2}), the forcing term is not so simple. Apart from the trigonometric and exponential terms, the Faddeeva function as well as functions $\Isin$ and $\Icos$ also appear in it.
Consequently, the Laplace transform method cannot be directly applied, as the inverse Laplace transform arising from this component cannot be expressed in terms of known functions. To circumvent this difficulty, we replace the Faddeeva, Isin and Icos functions by suitable approximations derived in Appendix \ref{apFunctions}. More exactly, only the terms representing a permanent oscillation have been retained. The complete analytical solutions of components $\Wt^{(2)}_1$ and $\Wt^{(2)}_2$ are given in appendixes \ref{apW2_1} and \ref{apW2_2} respectively.

Once the radius-weighted relative velocity is known, the actual bubble velocity is easily obtained, since the analytical solution of the external flow is also known. Indeed, for the forcing considered,
\begin{equation}
\tilde{v}_\infty (\tau)= \varepsilon M \Sto\, \sin(\Sto\, \tau )\left[ H(\tau) - H(\tau -\tau _p)\right],
\label{eq:dimensionless_acoustic_velocity_pulse_forcing}
\end{equation}
so the bubble velocity finally results:
\begin{equation}
\tilde{V}(\tau) = \varepsilon M \Sto \sin(\Sto\, \tau )\left[ H(\tau) - H(\tau -\tau _p)\right] - \varepsilon M \Sto\,\frac{\left(\Wt^{(1)}(\tau) + \varepsilon \Wt^{(2)}(\tau)\right)}{a(\tau)}.
\label{eq:leading_order_bubble_velocity}
\end{equation}

\section{Results and Discussion\label{sec:discussion}}

\subsection{Numerical results}

We compare here the bubble velocities predicted by the numerical solution of the full problem formulated in \S\ref{sec:problem_formulation} with those obtained from a simplified version of the system without the history integral. Moreover, we also employ the asymptotic approach presented in the previous section to evaluate its accuracy versus the full solution. For the sake of conciseness, three representative cases have been selected, whose conditions are reported in table \ref{table:cases}. They are all close to resonance, since it is when the bubble moves at the highest speed, which is the situation desired in practical applications. We would like to stress here that even though the number of cycles is relatively large in these cases ($\sim 100$), still memory effects are important to compute translation velocities. Were we using shorter pulses, like those used in medical imaging ($\sim 10$), these effects would be even more important.

\begin{figure}
\centering
\includegraphics[width=\textwidth]{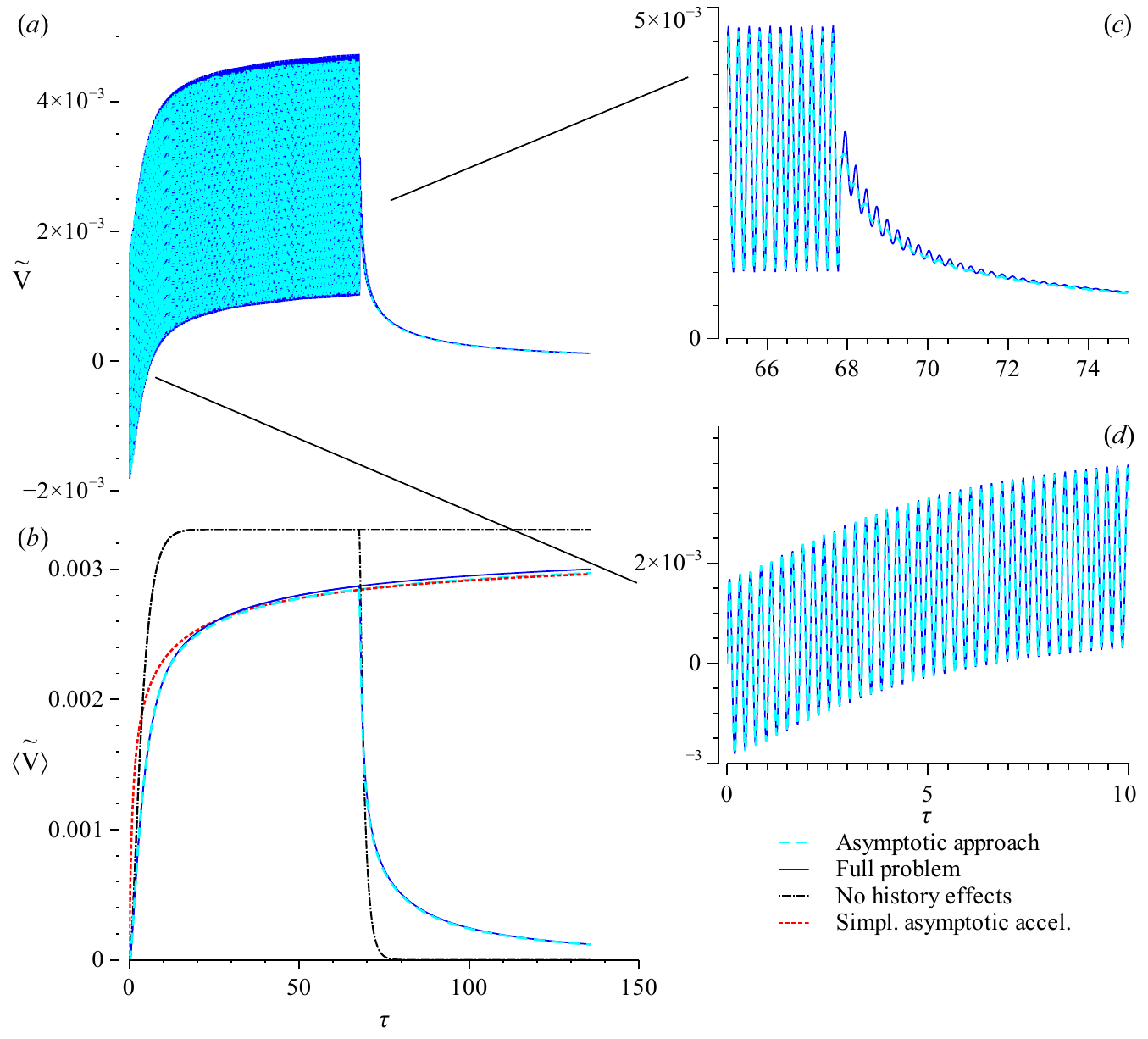}
\caption{\label{fig:case1}($a$) Comparison between the analytical solution (light blue dashed curve) and the numerical one (dark blue solid line) of the full problem for case I. ($b$) Period-averaged velocities. In this panel we also show the solution corresponding to the simplified problem without the memory term (black dashed-dotted line) and the simplified asymptotic solution for the acceleration process (red-dotted line), described in subsection \ref{sec:analytical_approximate}. From the point where the pulse stops ($\tau\approx 70$) onwards, two sets of curves are shown: one corresponds to the actual finite-duration pulse, whereas the other corresponds to a permanent insonation. Panels ($c$) and ($d$) show two magnifications of panel ($a$) corresponding to short and long times.}
\end{figure}

\begin{figure}
\centering
\includegraphics[width=\textwidth]{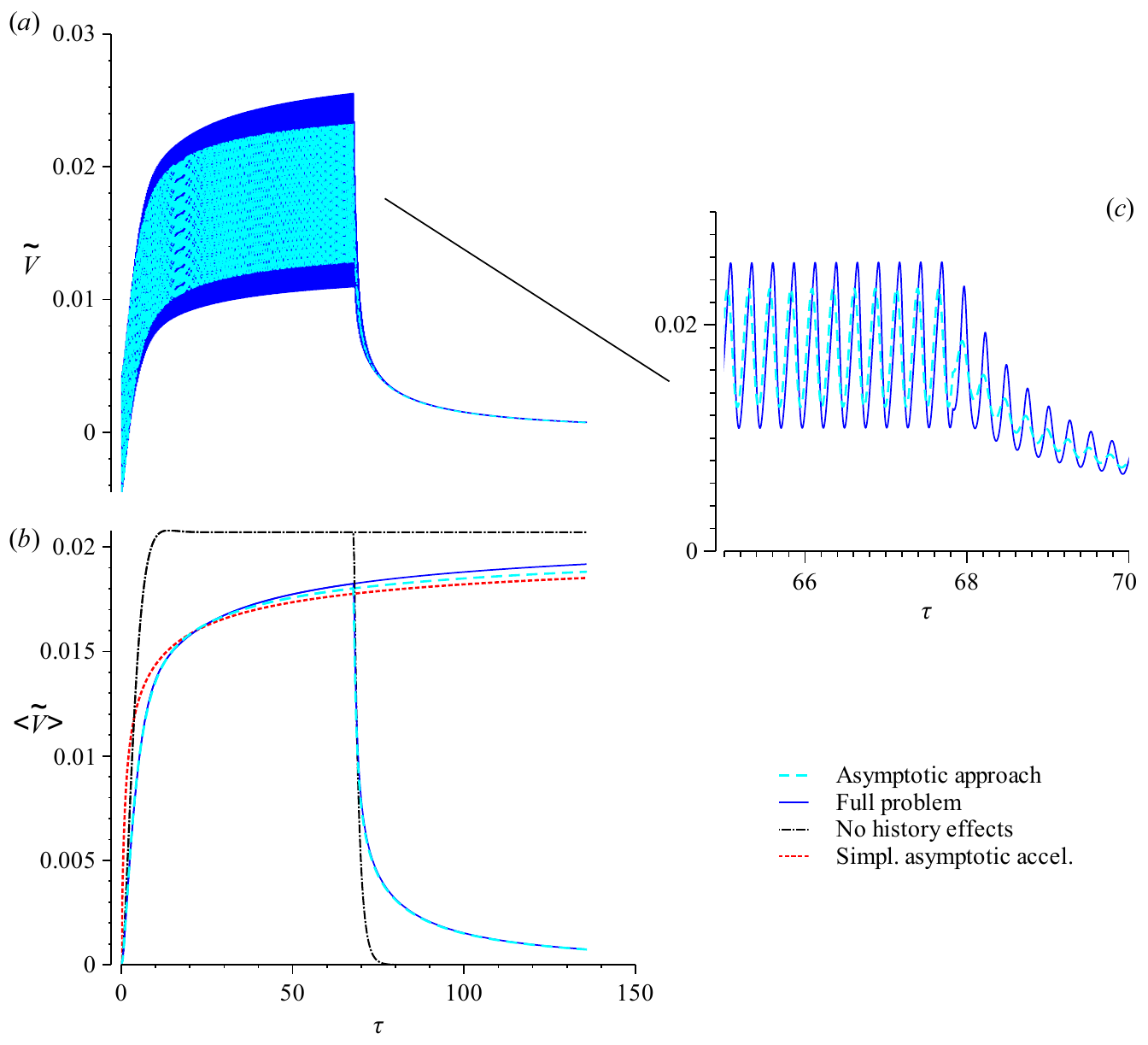}
\caption{\label{fig:case2}Comparison between the analytical solution and the numerical one of the full problem for case II. See caption of figure \ref{fig:case1} for an explanation of the different plots. Remarkably, although the amplitude of the oscillations is not well captured due to the relatively high amplitude of the forcing, still the mean velocity of the asymptotic solution agrees fairly well with the full numerical one.}
\end{figure}

\begin{figure}
\centering
\includegraphics[width=\textwidth]{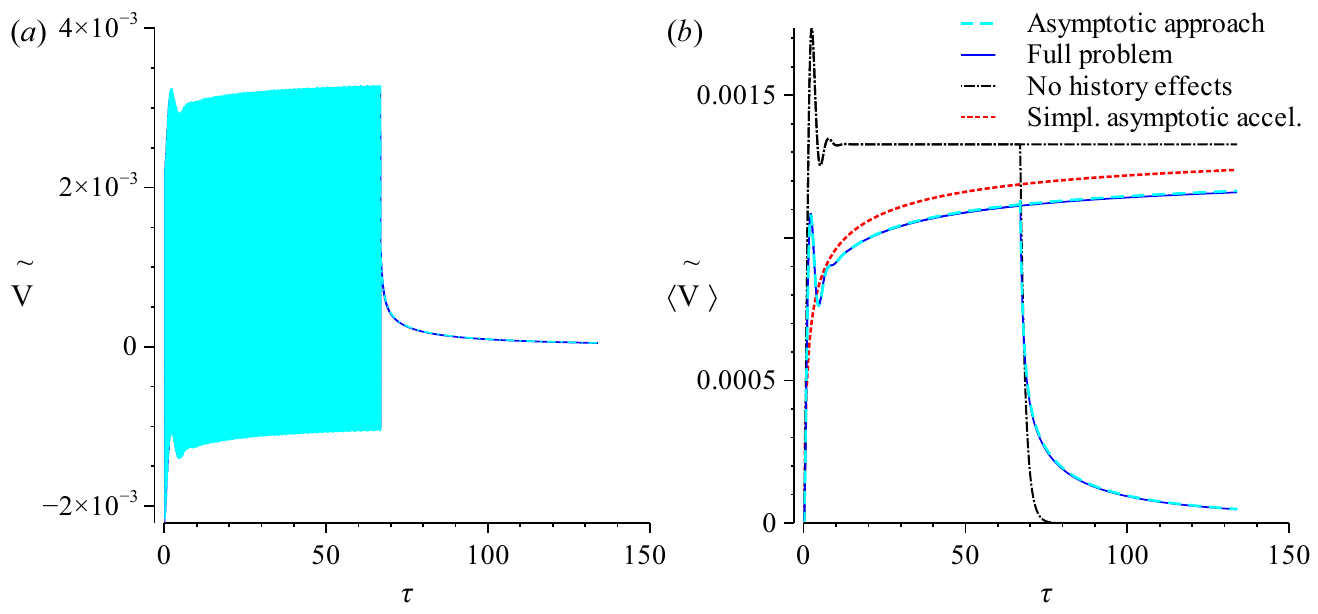}
\caption{\label{fig:case3}Comparison between the analytical solution and the numerical one of the full problem for case III. See caption of figure \ref{fig:case1} for an explanation of the different plots.}
\end{figure}

The results are plotted in figures \ref{fig:case1}, \ref{fig:case2} and \ref{fig:case3}. Panels labelled as ($a$) show the oscillatory velocities corresponding to the numerical solution of the problem (blue solid lines) and to the asymptotic solution (cyan dashed lines).
More interestingly, panels labelled as ($b$) contain the period-averaged velocities, defined as
\begin{equation}
\left<\tilde{V}\right> = \frac{\Sto}{2\pi}\int_\tau^{\tau+2\pi/\Sto} \tilde{V}(s)\,\rmd s.
\label{eq:period_average}
\end{equation}
In these panels, the period-averaged velocities obtained from the numerical solution of the problem without taking into account the history integral are also represented with black dot-dashed lines. As the experimental observations described in section \ref{sec:experimental_evidences} suggest, our calculations confirm that the proper description of the bubble velocity requires taking into account the memory integral, even though the pulses last for several tens of viscous times, i.e. a time much longer than the viscous one. Indeed, figures \ref{fig:case1}($b$), \ref{fig:case2}($b$), and \ref{fig:case3}($b$) show that the period-averaged velocity calculated neglecting history achieves the terminal value after about ten viscous times, a period much shorter than the duration of the pulse. On the contrary, the velocity predicted by the full solution approaches the asymptote much more slowly, not even reaching it while the acoustic excitation is on. In figures \ref{fig:case1}($b$), \ref{fig:case2}($b$), and \ref{fig:case3}($b$) the solutions have also been extended beyond the duration of the acoustic pulse, to illustrate the slow convergence towards the terminal velocity. A similar trend is observed once the pulse stops: the velocity does not decay exponentially, as would be expected if no memory integral was considered, but rather potentially.
\begin{table}
\centering
\begin{tabular}{ccccccccccc}
case & $\varepsilon$ & $\Sto$ & $\varepsilon\Sto$ & $M$ & $\Omega_0^{-1}$ & $R_0$ ($\mu$m) & $f$ (kHz) & $p_A$ (kPa) & No. cycles & $t_v$ (ms) \\
I & 0.0017 & 24.086 & 0.041 & 0.0146 & 0.9976 & 10 & 345 & 0.8 & 260 & 0.0479 \\
II & 0.0043 & 24.086 & 0.103 & 0.0146 & 0.9976 & 10 & 345 & 2.0 & 260 & 0.0479 \\
III & 0.001 & 50.0 & 0.05 & 0.0146 & 1.024 & 20.76 & 166.25 & 0.47 & 532 & 0.0111 \\
\end{tabular}
\caption{\label{table:cases}Dimensionless parameters and characteristic variables corresponding to the four cases considered. For all the cases $\sigma = 0.0728$ N/m, $c = 1485$ m/s and $\kappa = 1.4$.}
\end{table}

\begin{figure}
\centering
\includegraphics[width=\textwidth]{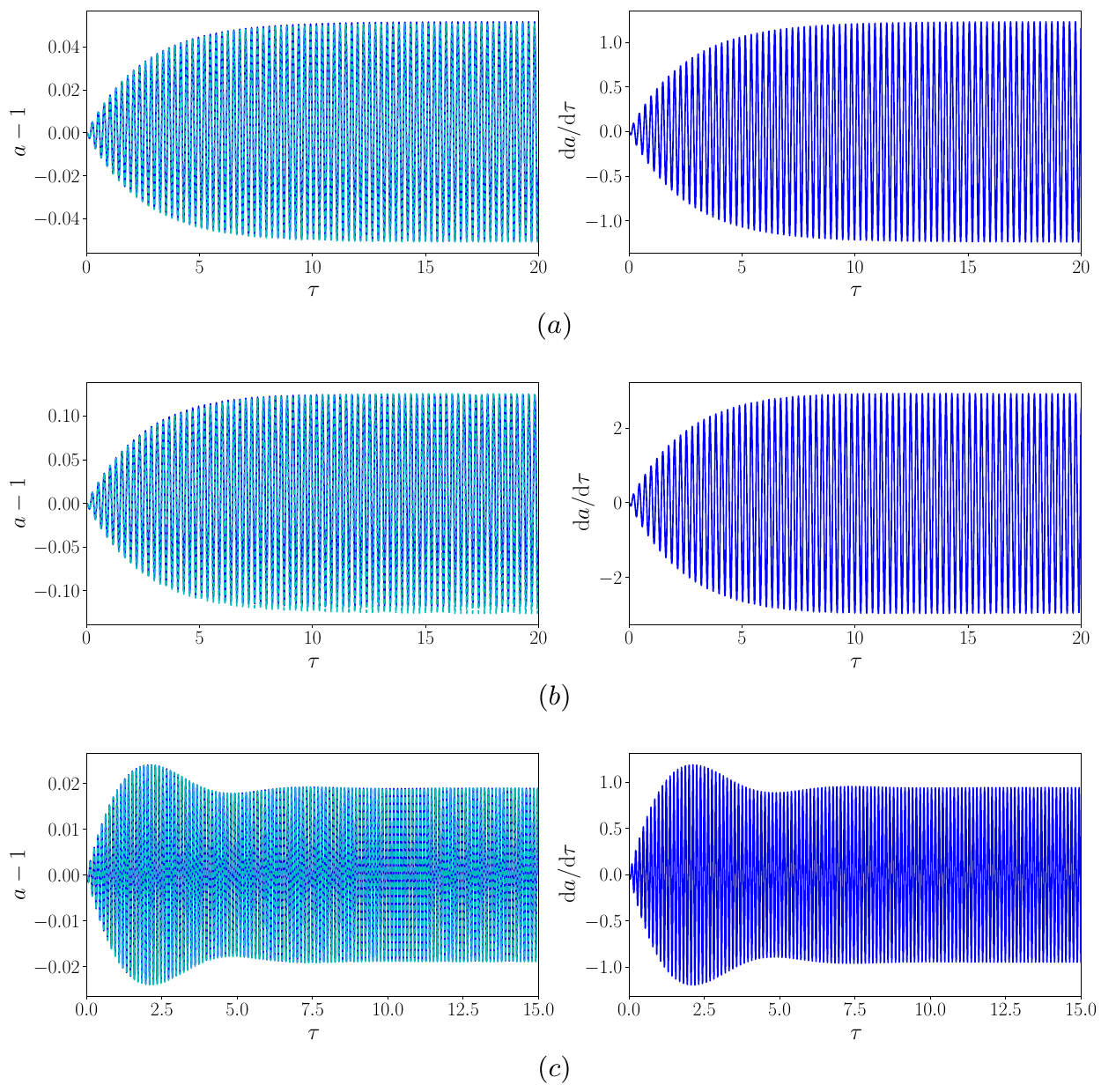}
\caption{\label{fig:radii}Time evolution of the radial oscillations and their time derivatives for the three cases detailed in table \ref{table:cases} (blue solid lines). The left panels also show the radial oscillation computed with the linearized equation (\ref{eq:linear_keller_miksis}) (cyan dashed lines).}
\end{figure}

For the sake of completeness, the first few cycles of the radial oscillations are also shown in figure \ref{fig:radii} for the three cases considered, along with their time derivative.
The behavior of the full numerical solution for the bubble velocity is fairly well captured by the asymptotic solution derived in subsection \ref{sec:leading_order}, specially that of the period-averaged velocity, which is the responsible for the net bubble translation.
This is quite remarkable taking into account that the cases reported in this work are all relatively close to the resonance frequency of the bubble. We have chosen cases close to resonance since they imply that the amplitude of the radial oscillations is going to be largest, which translates into strongest Bjerknes force, and thus to very large bubble displacements. In other words, this means that it is near resonance when the formulation developed here is going to be of greatest interest for the applications.
Although near resonance the amplitude of the radial oscillations may not be small in general, we impose an additional requirement to the parameters of the problem, namely $\delta \ll 1$, which translates into $\varepsilon\Sto \ll 1$ when the oscillations occur close to resonance (see Equation (\ref{eq:simplified_radius})).
It is worth paying attention to case II, where this condition is nearest to not being satisfied ($\varepsilon \Sto \sim 0.1$). It is interesting to observe how, even if the amplitude of the oscillations is not properly captured, still the average velocity is barely affected. This is consistent with the fact that, in the weakly non-linear regime, the response of the bubble differs from the linear solution in the appearance of harmonics and sub-harmonics that do not contribute to the average of the forcing terms in Equation (\ref{eq:first_order_eq}).

The simplified solution presented in subsection \ref{sec:analytical_approximate} (short-dashed red line in figures \ref{fig:case1}($b$), \ref{fig:case2}($b$), and \ref{fig:case3}($b$)) also describes fairly well the averaged solution of the full problem while the acoustic forcing is on. In particular, once the transient behavior in the bubble radial oscillations has damped out. Thus, although the exact evaluation of the bubble velocity taking into account history effects may be computationally expensive or analytically involved, the procedure followed in subsection \ref{sec:analytical_approximate} allows for an accurate description of the bubble dynamics while keeping mathematical or computational complexity within reasonable limits. Interestingly, it seems that to capture the transient bubble dynamics it suffices to retain the memory integral only at first order, whereas the transient volume oscillations and the history integral at leading order seem to influence the bubble velocity only at very short times.

It is reasonable to expect that this simplified solution may not describe well the dynamics of a bubble insonated with a pulse consisting of a very small number of cycles, in particular when the duration of the pulse is smaller or of the order of a few viscous times, where the volume oscillations have not yet reached the permanent regime. Notice that this limitation only affects the simplified solution described in \S\ref{sec:analytical_approximate}, not the more rigorous asymptotic approach presented in \S\ref{sec:leading_order}, which properly accounts for the effect of the transient volume oscillations. It should be pointed out that numerical simulations performed for the same conditions reported in table \ref{table:cases} but with a duration of ten pulses reveal that the history force is still very important and that the solution with no history term completely overpredicts the bubble velocity after the very first cycles.

Moreover, the good agreement between the numerical results and the simplified solution suggests that it can be used to infer how the bubble velocity approaches the steady translational velocity. To that end, we introduce into (\ref{eq:simplified_relative_velocity}) the asymptotic approximations of Appendix \ref{apFunctions}, which results into
\begin{equation}
\tilde{V}_{no} \approx -\varepsilon^2 M\Sto A_F\left(F_{s1} + \frac{C_{s1}/\sqrt{s_1} - 8D_{s2}/3}{\sqrt{\pi\tau}}\right),
\end{equation}
neglecting terms that decay exponentially. Consequently, we can conclude that the bubble net translational velocity approaches the steady one as $\tau^{-1/2}$ rather than exponentially, as would be the case were the memory effects neglected. This explains the slow convergence observed in our numerical simulations and supports the conclusion that history is indeed very important in order to accurately compute the bubble dynamics in many practical situations.

Finally, we would like to point out that most of the conclusions mentioned here concerning the rate at which the velocity converges to an asymptote are of application for bubbles in a fluid contaminated with surfactants. Indeed, in such case, the kernel that must be used in the memory integral is that of a solid sphere, namely $G(\tau-\overline{\tau}) = \left(\pi\left(\tau-\overline{\tau}\right)\right)^{-1/2}$. Using the approximations of appendix \ref{apFunctions}, it can be shown that the kernel for a clean bubble converges to this one at long times.

\subsection{\label{sec:comparison_experiments}Comparison with experiments}

We come back now to the experimental results that motivated this study. We compare the experimental deceleration of the bubble after the pulse (see section \ref{sec:experimental_evidences}) with the predictions obtained using the two extreme values of the Stokes' coefficients ($4\pi$ and $12\pi$) and with the analytical solution considering the effect of the history force.

We focus here on the deceleration period, as a direct comparison of the bubble dynamics during the pulse is not feasible at the large pressure amplitudes used in the experiments. In fact, the radial dynamics at these pressures are not well predicted by any equation that assumes radial symmetry, as shape oscillations may appear. However, these high pressures are needed in our experiments for the displacements to be large enough to be accurately measured. Thus, we compare instead the asymptotic trend that the bubble velocity follows at long times upon the pulse is off with the experimental results.

It can be easily seen that, in absence of forcing, the time evolution of the bubble velocity is given by Equation (\ref{eq:U0_analytic}), but replacing $\left.\Wt^{(1)}\right|_0$ by $\left.\tilde{V}\right|_0$, namely the velocity at the moment the pulse ends:
\begin{multline}
\tilde{V} (\tau) =   \tilde{V}|_0 \left\{ \rme^{-\alpha\tau}\left( A_0\cos(\gamma\tau)+
\frac{B_0-A_0\alpha}{\gamma}\sin(\gamma\tau) \right) 
 + C_0\rme^{s_1\tau}\erfc\sqrt{s_1 \tau}+\right. \\
 \left.+\frac{4}{3} \left[ \left( E_0 - 2\alpha D_0 \right)\Icos(\tau, \alpha, \gamma)
 + \frac{-E_0\alpha + D_0(\alpha^2-\gamma^2)}{\gamma}\Isin(\tau, \alpha, \gamma) \right]\right\},
 \label{v_decel_analytic}
\end{multline}
 \begin{figure}
 \centering
 \includegraphics[width=0.75\textwidth]{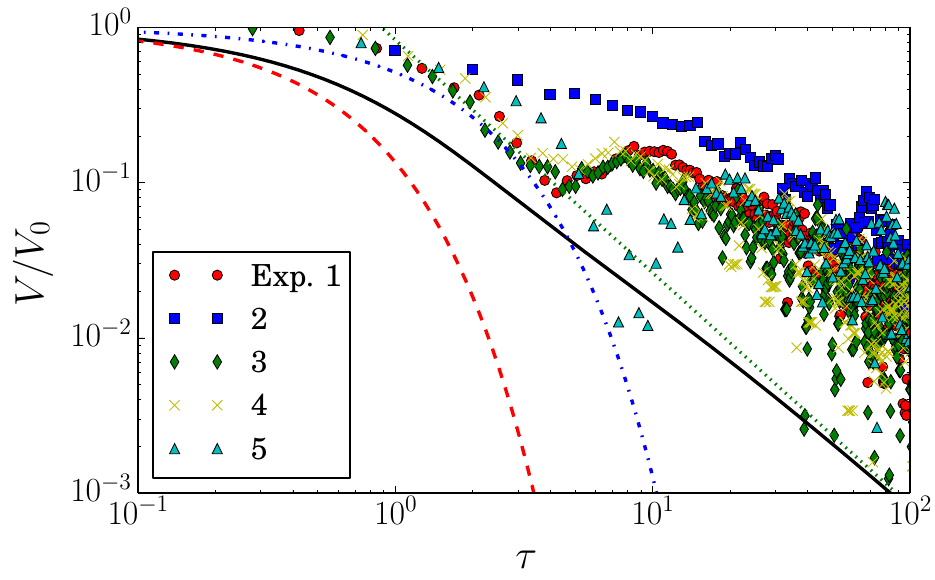}
 \caption{Time evolution of the velocity after the acoustic pulse for the experiments reported in table \ref{table_exp_bubbles} (symbols). We compare them with the velocities predicted by Equation (\ref{eq:stokes_drag}) using a constant Stokes' drag coefficient $C_d = 4\pi$ (dash-dotted blue line) and a coefficient $C_d = 12\pi$ (dashed red line), as well as with the analytical expressions taking into account memory effects: Equation (\ref{v_decel_analytic}) (solid gray line) and the simplified formula valid for long times, Equation (\ref{v_decel_simplified}) (green-dotted line), described in subsection \ref{sec:analytical_approximate}.}
 \label{fig:Deceleration_comparison}
  \end{figure}
Figure \ref{fig:Deceleration_comparison} shows that this solution, considering the history term for $Re = 0$, describes the long-time trend of the evolution of the bubble velocity much better than assuming a constant Stokes' coefficient. Still, the precise value of the velocity is not well captured. This is a consequence of the large radial oscillations that the bubble is expected to undergo after the finalization of the acoustic pulse, which modify the initial conditions at the beginning of the deceleration stage. It should be pointed out that, at short times, the curve corresponding to a constant coefficient $C_d = 4\pi$ seems to fit better the data. However we must keep in mind that, firstly, the bubble is still undergoing strong radial oscillations at these times, as is made evident by the rebound that occurs at $\tau \approx 10$, and secondly, that this curve should be precisely the behavior expected at long times if there was no memory force, being the curve corresponding to $C_d = 12\pi$ the expected behavior during the early stage \citep{MagnaudetLegendrePoF1998}. For this reason we find reasonable to claim that the curve obtained from our analytical solution (\ref{v_decel_analytic}) is a better description of the whole time range observed in the experiment.

Finally, notice that Equation (\ref{v_decel_analytic}) may be simplified in the limit $\tau \rightarrow \infty$, by making use again of the approximations in Appendix \ref{apFunctions}, to yield
\begin{equation}
\frac{\tilde{V}(\tau)}{\left.\tilde{V}\right|_0} \approx \frac{1}{2\sqrt{\pi}\tau^{3/2}}\left[ -\frac{C_0}{s_1^{3/2}}-\frac{4}{3}\frac{E_0}{\alpha^2+\gamma^2}
+ \frac{2}{3}\frac{D_0 \alpha\left(\alpha^2-\gamma^2\right)}{\left( \alpha^2+\gamma^2\right)^2} \right].
\label{v_decel_simplified}
\end{equation}
It can be observed in figure \ref{fig:Deceleration_comparison} that this approximation converges to the asymptotic solution in times of the order of $50$ viscous times. This long-time behavior reveals that the rate of deceleration of the bubble is potential in time, instead of exponential.

\section{Conclusions\label{sec:conclusions}}

The problem of the low-Reynolds-number translation of a clean bubble under the action of a finite-duration acoustic pulse is considered. A consistent formulation of the problem has been derived based mainly on the work of \cite{MagnaudetLegendrePoF1998}. The main feature of the theory presented here with respect to other similar approaches, most notably that presented by \cite{ToilliezSzeriJASA2008}, is that the present methodology can be applied to acoustic pulses of finite duration and thus allows us to study transient effects. Therefore, the memory integral term, commonly neglected in the literature, must be considered. Far from being a minor contribution, we have shown that this term can significantly affect the velocity achieved by the bubble in finite-duration acoustic pulses. We believe that this result is of interest for applications where the duration of the insonation is limited, such as medical applications or bubble control in artificial photosynthesis. Indeed, in many situations, the insonation is not continuous, but consists in a train of pulses whose duration may not be much longer than the bubble's viscous time. 

As a particular example of application of the methodology, we have derived an asymptotic solution for the case of a sinusoidal traveling wave modulated by a constant, but small, amplitude. The smallness of the amplitude reduces the full nonlinear problem to a set of three linear equations: one for the radial oscillations and two more for the leading order and the first order correction of the bubble relative velocity, respectively. This solution has proven to reproduce fairly well the numerical solution of the full problem even for cases close to resonance, where the amplitude of the radial oscillation is not so small.

Since the asymptotic solution is complicated to evaluate and understand due to its length, we also provide the reader with a simplified solution that still agrees fairly well with the full numerical results during the bubble acceleration period. In few words, this solution is obtained by neglecting transient effects in the bubble volume oscillations as well as in the leading order dynamic equation, thus only retaining the memory integral at first order, where a net force appears.

Besides the interest for the particular application of a bubble translating under the action of a traveling acoustic wave, it is worth noticing that this theory could be extended to describe other scenarios such as that of a bubble oscillating in a standing wave. Indeed, there are evidences that the memory integral term is of great importance even if the forcing is continuous. For instance, \cite{ToegelLutherLohsePRL2006} found out that incorporating the history force in the equations describing the translation of a bubble in sonoluminesce experiments, even in an approximate way, is crucial to explain the instabilities observed in their motion.

To conclude, we would like to point out some further limitations of our model. Firstly, in this work we have ignored the possibility of bubbles departing from the spherical shape. However, if the theory is to be applied for forcing frequencies close to the resonance ones, this possibility must be considered. This is specially true in the case of ultrasound contrast agents, due to the large volume oscillations that these bubbles experience \citep{Dollet_etalUMB2008}. In this sense, a modification of the theory to account for the translation of nearly spherical bubbles such as that developed by \cite{Romero_etalPoF2014} could be of interest. Secondly, we have not considered that bubbles may have some sort of coating or be covered by surfactants. A simple modification of our theory to account for the effect of non-clean bubble surface could be done by simply modifying the surface tension and liquid viscosity \citep{Faez_etalIEEE2013} on the one hand and replacing the shear-free boundary condition by a non-slip one on the other hand. However, these modifications are not expected to affect the main conclusions of our work.

We acknowledge the support of the Spanish Ministry of Economy and Competitiveness through grants DPI2014-59292-C3-1-P and DPI2015-71901-REDT, partly funded through European Funds. We would like to thank the reviewers for their comments and suggestions, specially one who has done a very extensive check of our calculations. 

\appendix

%
%
\section{Definition of functions and their approximation}
\label{apFunctions}
Functions $\Icos(\tau,a,b)$, $\Isin(\tau,a,b)$ and the so-called Faddeeva function $\F$, which appear in the analytical solution of the bubble velocity, are defined below
\begin{equation}
\Icos(\tau,a,b)= \frac{1}{\sqrt{\pi}}\int\limits_{0}^{\tau}\frac{\rme^{-at}\cos(bt)}{\sqrt{\tau-t}}dt= \Real \left\{ \frac{\mathrm{exp}\left[ -(a-ib)\tau \right]
\mathrm{erf}\left[ \sqrt{-(a-ib)\tau} \right]}{\sqrt{-(a-ib)}} \right\}
\label{eq:Icos_complex}
\end{equation}
\begin{equation}
\Isin(\tau,a,b)= \frac{1}{\sqrt{\pi}}\int\limits_{0}^{\tau}\frac{\rme^{-at}\sin(bt)}{\sqrt{\tau-t}}dt= \Imag \left\{ \frac{\mathrm{exp}\left[ -(a-ib)\tau \right]
\mathrm{erf}\left[ \sqrt{-(a-ib)\tau} \right]}{\sqrt{-(a-ib)}} \right\}
\label{eq:Isin_complex}
\end{equation}
\begin{equation}
\F (\tau)= e^{s_1\,\tau}\erfc\left(\sqrt{s_1\,\tau} \right).
\end{equation}
As explained in subsection \ref{sec:leading_order}, in order to solve the first-order dynamic equation by the Laplace transform method, it is convenient to replace these functions by suitable approximations in the following way: on the one hand, the asymptotic behaviour of the Faddeeva function can be estimated with help of the approximation of the error function offered in section 7.1.3, \cite{Abramowitz}.
\begin{equation}
\erf(\sqrt{z}) \approx 1- \frac{e^{-z}}{\sqrt{\pi\,z}}\left(1 - \frac{1}{2z}\right), \medskip \left|z\right|\rightarrow \infty.
\label{eq:approx_erf}
\end{equation}
According to \ref{eq:approx_erf}, our Faddeeva function $\F$ may be approximated by 
\begin{equation}
\F (\tau) \approx \frac{1}{\sqrt{\pi s_1 \tau}}\left(1-\frac{1}{2 s_1 \tau}\right).
\label{eq:approx_Faddeeva}
\end{equation}
On the other hand, the asymptotic behavior of functions $\Isin$ and $\Icos$ can be similarly obtained by introducing (\ref{eq:approx_erf}) into Equations (\ref{eq:Icos_complex}) and (\ref{eq:Isin_complex}), leading to the expressions below in the particular case of $\Icos(\tau, \alpha, \gamma)$ and $\Isin(\tau, \alpha, \gamma)$:
\begin{equation}
\Icos(\tau, \alpha, \gamma) \approx \frac{\alpha}{\alpha^2+\gamma^2}\frac{1}{\sqrt{\pi\tau}} + \frac{\alpha^2-\gamma^2}{\left(\alpha^2+\gamma^2\right)^2}\frac{1}{2\sqrt{\pi}\tau^{3/2}}
\end{equation}
\begin{equation}
\Isin(\tau, \alpha, \gamma) \approx \frac{\gamma}{\alpha^2+\gamma^2}\frac{1}{\sqrt{\pi\tau}} + \frac{2\alpha\gamma}{\left(\alpha^2+\gamma^2\right)^2}\frac{1}{2\sqrt{\pi}\tau^{3/2}}.
\end{equation}
Finally, for functions $\Icos(\tau,0,\Sto)$ and $\Isin(\tau,0,\Sto)$:
\begin{equation}
\Icos(\tau,0,\Sto) \approx \frac{1}{\sqrt{2\Sto}}\left( \sin\left(\Sto\,\tau\right) + \cos\left(\Sto\tau\right)\right).
\label{eq:approx_Icos}
\end{equation}
\begin{equation}
\Isin(\tau,0,\Sto) \approx \frac{1}{\sqrt{2\Sto}}\left( \sin\left(\Sto\,\tau\right) - \cos\left(\Sto\tau\right)\right) + \frac{1}{\sqrt{\pi\,\Sto^2\,\tau}}
\label{eq:approx_Isin}
\end{equation}
Notice that in this case only terms up to $\tau^{-1/2}$ have been considered, as the next order, $\tau^{-3/2}$, is not used in the paper.

%
%
\section{Numerical constants involved in the simplified solution of subsection \ref{sec:analytical_approximate}}
\label{apSimplifiedSolution}
Constants $s_1, \alpha, \beta, \gamma$ naturally appear in the analytic solution of the velocity when applying the Laplace transform method. Therefore, they will be used throughout the rest of the appendixes.
Their definition is:
\begin{subeqnarray}
 s_1   = & \frac{1}{3} \left( q^{\frac{1}{3}} + q^{-\frac{1}{3}} \right)-1\\
\alpha = & 2\frac{7q^{\frac{2}{3}} -2q^{\frac{1}{3}} -2q }{\left( q^{\frac{2}{3}}-3q^{\frac{1}{3}} +1 \right)^2}\\
\beta  = & \sqrt{\frac{4}{3} \frac{q^{\frac{1}{3}}}{q^{\frac{2}{3}}-3q^{\frac{1}{3}}+1}}\\
\gamma = & \sqrt{\beta^2-\alpha^2}
\end{subeqnarray}
with $q=15+4\sqrt{14}$.

Moreover, in subsection \ref{sec:analytical_approximate}, the following constants are used: $C_{s1,2}$, $D_{s1,2}$, $E_{s1,2}$ and $F_{s1,2}$. They arise from the solution of a linear system whose system matrix is:
\begin{equation}
\renewcommand{\arraystretch}{1.3}
M_a = \left[\begin{array}{cccc}
  1 & 1 & 0 & 1 \\
  2\alpha & -s_1 & 1 & -s_1 + 2\alpha \\
  \beta^2 & 0 & -s_1 & \beta^2 - 2\alpha s_1 \\
  0 & 0 & 0 & -\beta^2 s_1 
 \end{array}\right].
\end{equation}
In particular, from the solution of these systems:
\begin{equation}
\renewcommand{\arraystretch}{1.3}
\left[\begin{array}{c}
C_{s1}\\
D_{s1}\\
E_{s1}\\
F_{s1}
\end{array}
\right]
 =
M_a^{-1} 
\left[\begin{array}{c}
 0\\
 2\\
 2\\
 -\frac{4}{3}
  \end{array}\right]
 \end{equation}
\begin{equation}
\renewcommand{\arraystretch}{1.3}
\left[\begin{array}{c}
C_{s2}\\
D_{s2}\\
E_{s2}\\
F_{s2}
\end{array}
\right]
 =
M_a^{-1} 
\left[\begin{array}{c}
 0\\
 0\\
 0\\
 1
  \end{array}\right]
 \end{equation}

%
%
\section{Analytical solution of the linearized bubble radius $r$}
\label{apBubbleRadius}
Assuming small-amplitude volume oscillations of the bubble, the dimensionless bubble radius can be rewritten as in Equation \ref{eq:expansion_a}, i.e.
\begin{equation*}
a \approx 1+\varepsilon \, r(\tau).
\end{equation*}
Introducing this expansion in the dimensionless Keller-Miksis equation (\ref{eq:dimensionless_keller_miksis}), the first-order leads to Equation \ref{eq:linear_keller_miksis}, which can be written as 
\begin{equation}
 \alpha_r\ddot{r} + \beta_r\dot{r} + \Sto^2 \Omega_0^2 r = -\Sto^2 f\left(\xi\right),
\label{eq:dimensionless_keller_miksis_1order}
\end{equation}
with $\alpha_r=1+4M/9\Sto$, $\beta_r=4/9+M\Sto\Omega_0^2$ and $\Omega_0^2 = 3\kappa\Pi_0 + \left(3\kappa-1\right)\We^{-1}$. Its analytic solution for a finite-duration pulse (\ref{eq:Pulse_function}) reads
\begin{align}
r(\tau)= -\frac{\Sto^3}{\alpha_r}\left\{ a_r\,\cos(\Sto\,\tau) + \frac{b_r}{\Sto}\sin(\Sto\,\tau) + 
 \rme^{-\delta\tau}\left( c_r\,\cos(\Omega\tau) + \frac{d_r-\delta\,c_r}{\Omega}\sin(\Omega\tau) \right) \right\} \nonumber \\
 + H(\tau-\tau_p)\left\{ C_{1r}\cos(\Sto(\tau-\tau_p)) + C_{2r}\sin(\Sto(\tau-\tau_p)) \nonumber \right.\\
\left. + \rme^{-\delta(\tau-\tau_p)} \left( C_{3r}\cos(\Omega(\tau-\tau_p)) + C_{4r}\sin(\Omega(\tau-\tau_p))  \right)
   \right\},
\end{align}
where $\tau_p$ is the dimensionless duration of the acoustic pulse and parameters $\delta$ and $\Omega$ are defined as
\begin{eqnarray}
\delta & = & \frac{\beta_r}{2\alpha_r}, \\
\Omega & = & \frac{\Omega_0\Sto}{\alpha_r}\sqrt{\alpha_r - \frac{\beta_r^2}{4\Omega_0^2\Sto^2}}.
\end{eqnarray}
Finally, $C_{ir}$ are numerical constants given by
\begin{align*}
C_{1r} = \frac{e_r\,\sin(\Sto\,\tau_p) + a_r\Sto\cos(\Sto\tau_p)}{\alpha_r} \nonumber \\
C_{2r} = \frac{\frac{f_r}{\Sto}\sin(\Sto\,\tau_p) + b_r\cos(\Sto\tau_p)}{\alpha_r} \nonumber \\
C_{3r} = \frac{q_r\sin(\Sto\,\tau_p) + c_r\,\Sto\cos(\Sto\tau_p)}{\alpha_r} \nonumber \\
C_{4r} = \frac{k_r-q_r\delta}{\alpha_r\Omega}\sin(\Sto\,\tau_p) +  \frac{d_r-c_r\delta}{\alpha_r\Omega}\Sto\cos(\Sto\,\tau_p)
\end{align*}
and constants $a_r, b_r, c_r, \, ...\, k_r$ arise from solving the following linear systems
\begin{equation}  
M_r =
\renewcommand{\arraystretch}{1.3}
\left[\begin{array}{ccccc}
  1 & 0 & 1 & 0 \\
  2\delta & 1 &  0 & 1 \\
  \Omega^2 + \delta^2 & 2\delta & 1 & 0 \\
  0 & \Omega^2 + \delta^2 & 0 & 1
  \end{array}\right]
 \end{equation}    
 \begin{equation}
\left[\begin{array}{c}
a_r\\
b_r\\
c_r\\
d_r
\end{array}\right]
 = M_r^{-1} \times
\left[\begin{array}{c}
 0\\
 0\\
 0\\
 1 
  \end{array}\right]
 , \; 
\left[\begin{array}{c}
 e_r\\
 f_r\\
 q_r\\
 k_r
  \end{array}\right]
  = M_r^{-1} \times
\left[\begin{array}{c}
   0\\
   0\\
   1\\
   0
    \end{array}\right].
 \end{equation}
 
%
%
\section{Analytical solution of $\Wt^{(1)}_f (\tau)$}
\label{apW1}

As introduced in Equation (\ref{eq:leading_order_eq_laplace}), the leading order term of the radius-weighted relative velocity, $\Wt^{(1)}$, is split into two components: one due to any initial value that this component may have at the beginning of the acceleration or deceleration periods and another one due to the forcing. Mathematically,
\begin{equation}
\Wt^{(1)}(\tau) =  \Wt^{(1)}_0 (\tau)+ \Wt^{(1)}_f (\tau),
\end{equation}
with
\begin{align}
\Wt^{(1)}_0 (\tau) =\Wt^{(1)}|_0 \left\{ \rme^{-\alpha\tau}\left( A_0\cos(\gamma\tau)+
\frac{B_0-A_0\alpha}{\gamma}\sin(\gamma\tau) \right) 
 + C_0\rme^{s_1\tau}\erfc\left[ \sqrt{s_1 \tau} \right] \nonumber \right.\\
 \left. +\frac{4}{3}\left[ \left( E_0 - 2\alpha D_0 \right)\Icos(\tau, \alpha, \gamma) +
 \frac{-E_0\alpha + D_0(\alpha^2-\gamma^2)}{\gamma}\Isin(\tau, \alpha, \gamma)  \right]   \right\}
  \label{eq:U0_analytic}
\end{align}
and
\begin{equation}
\Wt^{(1)}_f (\tau) = {\cal L}^{-1}\left[ {\cal F}(s)\frac{2\Sto \left( s^2+s-\frac{2}{3} + \frac{4}{3}\sqrt{s} \right)}{\left( s-s_1 \right)\left[ (s+\alpha)^2 + \beta^2-\alpha^2 \right]} \right].
\label{eq:Uf_analytic}
\end{equation}
The oscillatory component of the relative velocity due to the acoustic forcing is given by the sum of a steady component, with superscript 's', and a non-steady component, with superscript 'ns',
\begin{equation}
\Wt^{(1)}_f (\tau) = \Wt^{(1)s}_{f}(\tau) + \Wt^{(1)ns}_{f}(\tau) ,
\end{equation}
where each component reads
\begin{equation}
\Wt^{(1)s}_f (\tau) = (-2\Sto)\Wt^{(1)}_{f}(\tau; 0) 
\end{equation}
\begin{equation}
\Wt^{(1)ns}_f (\tau) =  (2\Sto)\Wt^{(1)}_{f}(\tau; \tau_p)
\end{equation}
The general component $\Wt^{(1)}_{f}(\tau; \, \tau_0)$ is defined below
\begin{align}
\Wt^{(1)}_{f}(\tau; \, \tau_0) = \,H\left(\tau-\tau_0\right)
\left \{  A_1\cos{(\Sto\left( \tau-\tau_0 \right))} 
+   \frac{B_1}{\Sto}\sin{(\Sto\left( \tau-\tau_0 \right))}  \nonumber \right.\\
\left. + C_1\,\rme^{s_1\left(\tau-\tau_0 \right)}\erfc
\left( \sqrt{s_1\left( \tau-\tau_0 \right)} \right) \nonumber \right.\\
\left. + \rme^{-\alpha(\tau-\tau_0)} \left(D_1\cos{(\gamma \left( \tau-\tau_0 \right))}
+ \frac{E_1-D_1\alpha}{\gamma}\sin{(\gamma \left( \tau-\tau_0 \right))} \right) \nonumber \right.\\
\left. -\frac{4}{3}F_1\,\Sto\,\Isin\left( \tau-\tau_0,0, \Sto \right)  
+ \frac{4}{3} J_1 \,\Icos\left( \tau-\tau_0,0, \Sto \right)  \nonumber \right.\\
+\left. \frac{4}{3}\left( K_1-2Q_1\alpha \right) \Icos\left( \tau-\tau_0,\alpha, \gamma \right) \nonumber \right.\\
\left. +\frac{4}{3} \left( \frac{-K_1\alpha + Q_1(\alpha^2-\gamma^2)}{\gamma}\right)\Isin \left( \tau-\tau_0,\alpha, \gamma \right)  \right \},
\end{align}
being $\tau_0$ the starting time of the corresponding component, i.e. $\tau_0=0$ for the steady component and $\tau_0=\tau_p$ for the non-steady one..\\
Constants $A_0, B_0, C_0, D_0, E_0, J_0$, which are involved in $\Wt^{(1)}_0$, the component of $\Wt^{(1)}$ due to the initial velocity, can be obtained 
by solving the following linear systems 
\begin{equation}
\renewcommand{\arraystretch}{1.3}
\left[\begin{array}{c}
A_0\\
B_0\\
C_0\\
\end{array}
\right]
 =
\left[\begin{array}{ccc}
  1 & 0 & 1 \\
  -s_1 & 1 &  2\alpha \\
  0 & -s_1 & \beta^2 
 \end{array}\right]  
 ^{-1} 
\left[\begin{array}{c}
 1\\
 1\\
 -\frac{2}{3}
  \end{array}\right]
 \end{equation}
\begin{equation}
\renewcommand{\arraystretch}{1.3}
\left[\begin{array}{c}
D_0\\
E_0\\
J_0\\
 \end{array}\right]
 =
\left[\begin{array}{ccc}
  1 & 0 & 1 \\
  -s_1 & 1 &  2\alpha \\
  0 & -s_1 & \beta^2 
  \end{array}\right]  
 ^{-1} 
\left[\begin{array}{c}
 0\\
 0\\
 1
  \end{array}\right]
\end{equation} 

Constants $A_1, B_1, C_1, D_1, E_1, F_1, J_1, L_1, K_1, Q_1$ are involved in $\Wt^{(1)}_f$, the component of $\Wt^{(1)}$ due to the acoustic forcing. Their value arise from solving the linear systems below
\begin{equation}  
M_f =
\renewcommand{\arraystretch}{1.3}
\left[\begin{array}{ccccc}
  1 & 0 & 1 & 1 & 0 \\
  2\alpha-s_1 & 1 &  2\alpha & -s_1 & 1 \\
  \beta^2-2\alpha s_1 & 2\alpha-s_1 & \beta^2 + St^2 & St^2 & -s_1 \\
  -s_1 \beta^2 & \beta^2 -2\alpha s_1 & 2\alpha St^2 & -St^2 s_1 & St^2 \\
  0 & -s_1 \beta^2 & \beta^2 St^2 & 0 & -s_1 St^2
  \end{array}\right]
 \end{equation}    
\begin{equation}
\renewcommand{\arraystretch}{1.3}
\left[\begin{array}{c}
A_1\\
B_1\\
C_1\\
D_1\\
E_1
 \end{array}\right]
= M_f^{-1} \times
\left[\begin{array}{c}
 0\\
 1\\
 1\\
-\frac{2}{3}\\
 0
  \end{array}\right]
 , \; 
\left[\begin{array}{c}
 F_1\\
 J_1\\
 L_1\\
 K_1\\
 Q_1
  \end{array}\right]
 = M_f^{-1} \times
\left[\begin{array}{c}
   0\\
   0\\
   0\\
   1\\
   0
    \end{array}\right]
\end{equation}  
%

%
%
\section{Analytical solution of $\Wt^{(2)}_1 (\tau)$}
\label{apW2_1}
The solution of the first-order dynamic equation $\Wt^{(2)} (\tau)$ is split into two components. The former, $\Wt^{(2)}_1 (\tau)$, due to the first forcing term reads
\begin{equation}
\Wt^{(2)}_1  (\tau) = \Wt^{(2)s}_{1}  (\tau) + \Wt^{(2)ns}_{1}  (\tau),
\end{equation}
where the superscript 's' means 'steady' whereas the superscript 'ns' means non-steady. \\
At the same time, these steady and non-steady components are given by
\begin{align}
\Wt^{(2)s}_{1}  (\tau) = \frac{2\Sto^2}{\alpha_r}\left[\sum_{j=1}^{3} \vWt^{(2)s}_{1,j} \left( \tau;\, 0, \, 0, \, 2\Sto \right) +
 \sum_{j=4}^{5} \Wt^{(2)s}_{1,j} \left( \tau;\, 0, \, \delta, \, \Omega + \Sto \right) \nonumber \right.\\
 \left. + \sum_{j=6}^{7} \Wt^{(2)s}_{1,j} \left( \tau;\, 0, \, \delta, \, \Omega - \Sto \right) \right]
\end{align}
\begin{align}
\Wt^{(2)ns}_{1}  (\tau) = -\frac{2\Sto^2}{\alpha_r}\left[\sum_{j=1}^{3} \Wt^{(2)ns}_{1,j} \left( \tau;\, \tau_p, \, 0, \, 2\Sto \right) +
 \sum_{j=4}^{5} \Wt^{(2)ns}_{1,j} \left( \tau;\, \tau_p, \, \delta, \, \Omega + \Sto \right) \nonumber \right.\\ 
 \left.  + \sum_{j=6}^{7} \Wt^{(2)ns}_{1,j} \left( \tau;\, \tau_p, \, \delta, \, \Omega - \Sto \right)\right],
\end{align}
where the general component $\Wt^{(2)s}_{i,j}(\tau; \,\tau_k, \, \varphi_1, \, \varphi_2)$, for both the steady and the non-steady case, reads
\begin{align}
\Wt^{(2)s}_{1,j}(\tau; \,\tau_k, \, \varphi_1, \, \varphi_2) = S^s_{1,j}\,H\left(\tau-\tau_k\right)
\left \{ A_{0,1,j} + C_{1,j}\rme^{s_1\left(\tau-\tau_k \right)}\erfc\left( \sqrt{s_1\left( \tau-\tau_k \right)} \right) \nonumber \right.\\
\left. + \rme^{-\varphi_1(\tau-\tau_k)} \left( A_{1,j}\cos{(\varphi_2\left( \tau-\tau_k \right))} 
+   B_{1,j}\,\sin{(\varphi_2\left( \tau-\tau_k \right))} \right)  \nonumber \right.\\
\left. + \rme^{-\alpha(\tau-\tau_k)} \left(D_{1,j}\cos{(\gamma \left( \tau-\tau_k \right))}
+ E_{1,j}\sin{(\gamma \left( \tau-\tau_k \right))} \right) \nonumber \right.\\
\left. +\frac{4}{3}F_{1,j}\,\Isin\left( \tau-\tau_k,\varphi_1, \varphi_2 \right)  
+ \frac{4}{3} J_{1,j} \,\Icos\left( \tau-\tau_k,\varphi_1, \varphi_2 \right)  \nonumber \right.\\
+\left. \frac{4}{3}K_{1,j}\, \Icos\left( \tau-\tau_k,\alpha, \gamma \right) +\frac{4}{3}Q_{1,j}\,\Isin \left( \tau-\tau_k,\alpha, \gamma \right)  \right \}.
\end{align}
Note that the constants involved in each component are given in Appendix \ref{apConstantsW2}.

%
%
\section{Analytical solution of $\Wt^{(2)}_2 (\tau)$}
\label{apW2_2}
The analytic solution of the component of $\Wt^{(2)} (\tau)$ due to the second forcing term, $\Wt^{(2)}_2 (\tau)$, is  
\begin{equation}
\Wt^{(2)}_2  (\tau) = \Wt^{(2)s}_{2,1}  (\tau) + \Wt^{(2)ns}_{2,1}  (\tau) + 
\Wt^{(2)s}_{2,2}  (\tau) + \Wt^{(2)ns}_{2,2}  (\tau),
\end{equation}
where the steady and non-steady components read
\begin{align}
\Wt^{(2)s}_{2,1}  (\tau) =\frac{-4\Sto^2}{\alpha_r} \left[ \sum_{k=1}^{3} \Wt^{(2)s}_{2,1,k} \left( \tau;\, 0, \, 0, \, 2\Sto \right) +
 \sum_{k=4}^{5} \Wt^{(2)s}_{2,1,k} \left( \tau;\, 0, \, \delta, \, \Omega + \Sto \right) \nonumber \right.\\ 
\left. +  \sum_{k=6}^{7} \Wt^{(2)s}_{2,1,k} \left( \tau;\, 0, \, \delta, \, \Omega - \Sto \right) \right]
\end{align}
\begin{align}
\Wt^{(2)ns}_{2,1}  (\tau) =-4\Sto \left[ \sum_{k=1}^{3} \Wt^{(2)ns}_{2,1,k} \left( \tau;\, \tau_p, \, 0, \, 2\Sto \right) +
 \sum_{k=4}^{5} \Wt^{(2)ns}_{2,1,k} \left( \tau;\, \tau_p, \, \delta, \, \Omega + \Sto \right) \nonumber \right.\\
 \left. +  \sum_{k=6}^{7} \Wt^{(2)ns}_{2,1,k} \left( \tau;\, \tau_p, \, \delta, \, \Omega - \Sto \right)\right]
\end{align}
\begin{align}
\Wt^{(2)s}_{2,2}  (\tau) = \frac{-4\Sto^2}{\alpha_r}\left[\sum_{k=1}^{2} \Wt^{(2)s}_{2,2,k} \left( \tau;\, 0, \, \alpha, \, \gamma + \Sto \right) +
 \sum_{k=3}^{4} \Wt^{(2)s}_{2,2,k} \left( \tau;\, 0, \, \alpha, \, \gamma - \Sto \right) \nonumber \right.\\ 
\left. +  \sum_{k=5}^{6} \Wt^{(2)s}_{2,2,k} \left( \tau;\, 0, \, \alpha+\delta, \, \Omega + \gamma \right) + 
  \sum_{k=7}^{8} \Wt^{(2)s}_{2,2,k} \left( \tau;\, 0, \, \alpha+\delta, \, \Omega - \gamma \right)\right]
\end{align}
\begin{align}
\Wt^{(2)ns}_{2,2}  (\tau) = -4\Sto \left[  \sum_{k=1}^{2} \Wt^{(2)ns}_{2,2,k} \left( \tau;\, \tau_p, \, \alpha, \, \gamma + \Sto \right) +
 \sum_{k=3}^{4} \Wt^{(2)ns}_{2,2,k} \left( \tau;\, \tau_p, \, \alpha, \, \gamma - \Sto \right) \nonumber \right.\\
\left. +  \sum_{k=5}^{6} \Wt^{(2)ns}_{2,2,k} \left( \tau;\, \tau_p, \, \alpha+\delta, \, \Omega + \gamma \right) + 
  \sum_{k=7}^{8} \Wt^{(2)ns}_{2,2,k} \left( \tau;\, \tau_p, \, \alpha+\delta, \, \Omega - \gamma \right)\right].
\end{align}
Finally, the general component $\Wt^{(2)s}_{i,j,k}(\tau; \,\tau_k, \, \varphi_1, \, \varphi_2)$, for both the steady and the non-steady case, is given by
\begin{align}
\Wt^{(2)s}_{2,j,k}(\tau; \,\tau_k, \, \varphi_1, \, \varphi_2) = S^s_{2,j,k}\,H\left(\tau-\tau_k\right)
\left \{ A^s_{2,j,k} + C^s_{2,j,k}\rme^{s_1\left(\tau-\tau_k \right)}\erfc\left( \sqrt{s_1\left( \tau-\tau_k \right)} \right) \nonumber \right.\\
\left. + \rme^{-\varphi_1(\tau-\tau_k)} \left( A^s_{2,j,k}\cos{(\varphi_2\left( \tau-\tau_k \right))} 
+   B^s_{2,j,k}\,\sin{(\varphi_2\left( \tau-\tau_k \right))} \right)  \nonumber \right.\\
\left. + \rme^{-\alpha(\tau-\tau_k)} \left(D^s_{2,j,k}\cos{(\gamma \left( \tau-\tau_k \right))}
+ E^s_{2,j,k}\sin{(\gamma \left( \tau-\tau_k \right))} \right) \nonumber \right.\\
\left. +\frac{4}{3}F^s_{2,j,k}\,\Isin\left( \tau-\tau_k,\varphi_1, \varphi_2 \right)  
+ \frac{4}{3} J^s_{2,j,k} \,\Icos\left( \tau-\tau_k,\varphi_1, \varphi_2 \right)  \nonumber \right.\\
+\left. \frac{4}{3}K^s_{2,j,k}\, \Icos\left( \tau-\tau_k,\alpha, \gamma \right) +\frac{4}{3}Q^s_{2,j,k}\,\Isin \left( \tau-\tau_k,\alpha, \gamma \right)  \right \}.
\end{align}
The constants involved in each component are given in Appendix \ref{apConstantsW2}. 

%
%
\section{Constants involved in the computation of the first-order correction of the relative velocity, $\Wt^{(2)}$}
\label{apConstantsW2}
\subsection{Constants for $\Wt^{(2)s}_{1,1}$ and $\Wt^{(2)ns}_{1,1}$ }
In general, constants $A_{1,j}, \, B_{1,j}, \, ... Q_{1,j}$ are the same for both the steady and the non-steady part.
Their values for the first component can be obtained through the resolution of the following linear system
\begin{equation*}  
M_{1,1} =
\renewcommand{\arraystretch}{1.3}
\left[\begin{array}{ccccc}
  1 & 1 & 1 & 0 \\
  2\alpha-s_1 & 2\alpha &  -s_1 & 1 \\
  \beta^2-2\alpha s_1 & \beta^2 & 0 & -s_1 \\
  -s_1\beta^2 & 0 & 0 & 0
  \end{array}\right]
 \end{equation*}    
 \begin{equation*}
\left[\begin{array}{c}
A_{1,1}\\
C_{1,1}\\
D_{1,1}\\
E_{1,1}
\end{array}\right]
 =(M_{1,1})^{-1} \times
\left[\begin{array}{c}
 0\\
 1\\
 1\\
 -2/3 
  \end{array}\right]
 , \; 
\left[\begin{array}{c}
 F_{1,1}\\
 H_{1,1}\\
 Q_{1,1}\\
 K_{1,1}
  \end{array}\right]
  =(M_{1,1})^{-1} \times
\left[\begin{array}{c}
   0\\
   0\\
   0\\
   1
    \end{array}\right].
 \end{equation*}
 \begin{eqnarray*}
 A_{0,1,1} & = & A_{1,1}\\ 
 B_{1,1} & = & G_{1,1} =0
 \end{eqnarray*}
 Moreover, constant $S^s_{1,1}$ is also equal to $S^{ns}_{1,1}$ in this case, 
 \begin{eqnarray*}
 S^s_{1,1} & = & S^{ns}_{1,1} = \frac{a_r}{2}
 \end{eqnarray*}
\subsection{Constants for $\Wt^{(2)s}_{1,2}$ and $\Wt^{(2)ns}_{1,2}$ }
\begin{equation*}  
M_{1,2}(\varphi) =
\renewcommand{\arraystretch}{1.3}
\left[\begin{array}{ccccc}
  1 & 0 & 1 & 1 & 0 \\
  2\alpha-s_1 & 1 & 2\alpha &  -s_1 & 1 \\
  \beta^2-2\alpha s_1 & 2\alpha-s_1 & \varphi^2+ \beta^2 & \varphi^2 & -s_1 \\
  -\beta^2s_1 & \beta^2- 2\alpha s_1 & 2\alpha \varphi^2 & - \varphi^2 s_1 & \varphi^2 \\
  0 & -\beta^2 s_1 & \varphi^2 \beta^2 & 0 & -\varphi^2 s_1
  \end{array}\right]
 \end{equation*}    
 \begin{equation*}
\left[\begin{array}{c}
A_{1,2}\\
B_{1,2}\\
C_{1,2}\\
D_{1,2}\\
E_{1,2}
\end{array}\right]
 = (M_{1,2}(2\Sto))^{-1}\times
\left[\begin{array}{c}
 0\\
 1\\
 1\\
 -2/3\\
 0
  \end{array}\right]
 , \; 
\left[\begin{array}{c}
 F_{1,2}\\
 G_{1,2}\\
 H_{1,2}\\
 Q_{1,2}\\
 K_{1,2}
  \end{array}\right]
  =(M_{1,2}(2\Sto))^{-1}\times
\left[\begin{array}{c}
   0\\
   0\\
   0\\
   1\\
   0
    \end{array}\right]
 \end{equation*}
with $\varphi=2\Sto$ and $A_{0,1,2}=0$.\\
Constants $S^s_{1,2}$ and $S^{ns}_{1,2}$ read
 \begin{eqnarray*}
 S^s_{1,2} & = & \frac{a_r}{2}  \\
 S^{ns}_{1,2} & = & \frac{b_r}{2\Sto}\sin(2\Sto \tau_p) + \frac{a_r}{2}\cos(2\Sto\tau_p)  
 \end{eqnarray*}
\subsection{Constants for $\Wt^{(2)s}_{1,3}$ and $\Wt^{(2)ns}_{1,3}$ }
 \begin{equation*}
 \left[\begin{array}{c}
 A_{1,3}\\
 B_{1,3}\\
 C_{1,3}\\
 D_{1,3}\\
 E_{1,3}
 \end{array}\right]
  = (M_{1,2}(2\Sto))^{-1}\times
 \left[\begin{array}{c}
  0\\
  0\\
  1\\
  1\\
  -2/3
   \end{array}\right]
  , \; 
 \left[\begin{array}{c}
  F_{1,3}\\
  G_{1,3}\\
  H_{1,3}\\
  Q_{1,3}\\
  K_{1,3}
   \end{array}\right]
   =(M_{1,2}(2\Sto))^{-1} \times
 \left[\begin{array}{c}
    0\\
    0\\
    0\\
    0\\
    1
     \end{array}\right].
  \end{equation*}
 \begin{eqnarray*}
 A_{0,1,3} & = & 0 \\
 S^s_{1,3} & = & b_r  \\
 S^{ns}_{1,3} & = & \left( \frac{b_r}{2\Sto}\cos(2\Sto\tau_p)- \frac{a_r}{2}\sin(2\Sto\tau_p) \right)2\Sto 
 \end{eqnarray*}
\subsection{Constants for $\Wt^{(2)s}_{1,4}$ and $\Wt^{(2)ns}_{1,4}$ }
\begin{eqnarray*}  
& M_{1,4}(\varphi_1, \varphi_2)= &\\
\renewcommand{\arraystretch}{1.3}
& \left[\begin{array}{ccccc}
  1 & 0 & 1 & 1 & 0 \\
  2\alpha-s_1 & 1 & 2(\alpha+\varphi_1) &  2\varphi_1-s_1 & 1 \\
  \beta^2-2\alpha s_1 & 2\alpha-s_1 & \varphi_1^2+\beta^2+4\alpha \varphi_1+ \varphi_2^2 & \varphi_1^2-2s_1\varphi_1 +\varphi_2^2 & 2\varphi_1-s_1 \\
 -s_1\beta^2 & \beta^2- 2\alpha s_1 & 2\alpha \varphi_1^2 + 2\alpha \varphi_2^2+2\varphi_1\beta^2 & -s_1(\varphi_1^2+\varphi_2^2) & \varphi_2^2+\varphi_1^2-2\varphi_1 s_1 \\
  0 & -\beta^2 s_1 & \beta^2(\varphi_1^2+\varphi_2^2) & 0 & -s_1(\varphi_1^2+\varphi_2^2)
  \end{array}\right],
 \end{eqnarray*}    
 \begin{equation*}
 \left[\begin{array}{c}
 A_{1,4}\\
 B_{1,4}\\
 C_{1,4}\\
 D_{1,4}\\
 E_{1,4}
 \end{array}\right]
  = (M_{1,4}(\delta, \Omega+\Sto))^{-1} \times
 \left[\begin{array}{c}
  0\\
  1\\
  1+\delta\\
  -2/3+\delta\\
  -2\delta/3
   \end{array}\right]
  , \; 
 \left[\begin{array}{c}
  F_{1,4}\\
  G_{1,4}\\
  H_{1,4}\\
  Q_{1,4}\\
  K_{1,4}
   \end{array}\right]
   =(M_{1,4}(\delta, \Omega+\Sto))^{-1} \times
 \left[\begin{array}{c}
    0\\
    0\\
    0\\
    1\\
    \delta
     \end{array}\right]
  \end{equation*}
with $\varphi_1=\delta$ and $\varphi_2=\Omega+\Sto$.\\
 \begin{eqnarray*}
  A_{0,1,4} & = & 0 \\
 S^s_{1,4} & = & \frac{c_r}{2}  \\
 S^{ns}_{1,4} & = & \left( \frac{c_r}{2}\cos\left((\Omega+\Sto)\tau_p\right) + 
 \frac{d_r-\delta c_r}{2\Omega}\sin\left((\Omega+\Sto)\tau_p\right) \right)\rme^{-\delta\tau_p}
 \end{eqnarray*}
\subsection{Constants for $\Wt^{(2)s}_{1,5}$ and $\Wt^{(2)ns}_{1,5}$ }
 \begin{equation*}
   \left[\begin{array}{c}
   A_{1,5}\\
   B_{1,5}\\
   C_{1,5}\\
   D_{1,5}\\
   E_{1,5}
   \end{array}\right]
    = (M_{1,4}(\delta, \Omega+\Sto))^{-1} \times
   \left[\begin{array}{c}
    0\\
    0\\
    1\\
    1\\
    -2/3
     \end{array}\right]
   , \;      
   \left[\begin{array}{c}
    F_{1,5}\\
    G_{1,5}\\
    H_{1,5}\\
    Q_{1,5}\\
    K_{1,5}
     \end{array}\right]
     =(M_{1,4}(\delta, \Omega+\Sto))^{-1} \times
   \left[\begin{array}{c}
      0\\
      0\\
      0\\
      0\\
      1
       \end{array}\right]
    \end{equation*}
 \begin{eqnarray*}
  A_{0,1,5} & = & 0 \\
 S^s_{1,5} & = & \frac{(d_r-c_r\delta)(\Omega+\Sto)}{2\Omega}  \\
 S^{ns}_{1,5} & = & \left( -\frac{c_r}{2}\sin\left((\Omega+\Sto)\tau_p\right) + 
 \frac{d_r-\delta c_r}{2\Omega}\cos\left((\Omega+\Sto)\tau_p\right) \right)\left( \Omega+\Sto\right)
 \rme^{-\delta\tau_p}
 \end{eqnarray*}
\subsection{Constants for $\Wt^{(2)s}_{1,6}$ and $\Wt^{(2)ns}_{1,6}$ }
 \begin{equation*}
  \left[\begin{array}{c}
  A_{1,6}\\
  B_{1,6}\\
  C_{1,6}\\
  D_{1,6}\\
  E_{1,6}
  \end{array}\right]
   = (M_{1,4}(\delta, \Omega-\Sto))^{-1} \times
  \left[\begin{array}{c}
   0\\
   1\\
   1+\delta\\
   -2/3+\delta\\
   -2\delta/3
    \end{array}\right]
   , \; 
  \left[\begin{array}{c}
   F_{1,6}\\
   G_{1,6}\\
   H_{1,6}\\
   Q_{1,6}\\
   K_{1,6}
    \end{array}\right]
    =(M_{1,4}(\delta, \Omega-\Sto))^{-1} \times
  \left[\begin{array}{c}
     0\\
     0\\
     0\\
     1\\
     \delta
      \end{array}\right]
   \end{equation*}
 \begin{eqnarray*}
  A_{0,1,6} & = & 0 \\
 S^s_{1,6} & = & \frac{c_r}{2}  \\
 S^{ns}_{1,6} & = & \left( \frac{c_r}{2}\cos\left((\Omega-\Sto)\tau_p\right) + 
 \frac{d_r-\delta c_r}{2\Omega}\sin\left((\Omega-\Sto)\tau_p\right) \right)\rme^{-\delta\tau_p}
 \end{eqnarray*}
\subsection{Constants for $\Wt^{(2)s}_{1,7}$ and $\Wt^{(2)ns}_{1,7}$ }
 \begin{equation*}
   \left[\begin{array}{c}
   A_{1,7}\\
   B_{1,7}\\
   C_{1,7}\\
   D_{1,7}\\
   E_{1,7}
   \end{array}\right]
    =(M_{1,4}(\delta, \Omega-\Sto))^{-1} \times
   \left[\begin{array}{c}
    0\\
    0\\
    1\\
    1\\
    -2/3
     \end{array}\right]
    , \; 
   \left[\begin{array}{c}
    F_{1,7}\\
    G_{1,7}\\
    H_{1,7}\\
    Q_{1,7}\\
    K_{1,7}
     \end{array}\right]
     =(M_{1,4}(\delta, \Omega-\Sto))^{-1} \times
   \left[\begin{array}{c}
      0\\
      0\\
      0\\
      0\\
      1
       \end{array}\right]
    \end{equation*}
 \begin{eqnarray*}
  A_{0,1,7} & = & 0 \\
 S^s_{1,7} & = & \frac{(d_r-c_r\delta)(\Omega-\Sto)}{2\Omega}  \\
 S^{ns}_{1,7} & = & \left( -\frac{c_r}{2}\sin\left((\Omega-\Sto)\tau_p\right) + 
 \frac{d_r-\delta c_r}{2\Omega}\cos\left((\Omega-\Sto)\tau_p\right) \right)\left( \Omega-\Sto\right)
 \rme^{-\delta\tau_p}
 \end{eqnarray*}
\subsection{Constants for $\Wt^{(2)s}_{2,1,1}$ and $\Wt^{(2)ns}_{2,1,1}$ }
In general, constants involved in this part of the first-order velocity coincide with the constants given in the previous subsection, i.e. $A_{1,k}=A_{2,1,k}$, $B_{1,k}=B_{2,1,k}$, $C_{1,k}=C_{2,1,k}$, ... However, constants $S^s_{2,1,k}$ and $S^{ns}_{2,1,k}$ are different for each component.\\
For simplicity, let's rename the following combination of constants as $Z_1$, $Z_2$,...,$Z_8$, which repeatedly appear in $S^s_{2,1,k}$ and $S^{ns}_{2,1,k}$:
 \begin{eqnarray*}
  Z_1 & = & Z_{11} \cos(\Sto\tau_p) - Z_{12}\sin(\Sto\tau_p)   \\
   Z_2 & = & Z_{11} \sin(\Sto\tau_p) + Z_{12}\cos(\Sto\tau_p) \\
   Z_3 & = & Z_{11} \cos(\Sto\tau_p) - Z_{12}\sin(\Sto\tau_p)- A_1 -\frac{4}{3}F_1\sqrt{\frac{\Sto}{2}}-\frac{4}{3}G_1\frac{1}{\sqrt{2\Sto}}\\
 Z_4 & = & Z_{11} \sin(\Sto\tau_p) + Z_{12}\cos(\Sto\tau_p) - \frac{B_1}{\Sto} + \frac{4}{3}F_1\sqrt{\frac{\Sto}{2}} - \frac{4}{3}G_1\frac{1}{\sqrt{2\Sto}} \\
  Z_5 & = & \rme^{\alpha\tau_p} \left( Z_{13} \cos(\gamma\tau_p) - Z_{14}\sin(\gamma\tau_p) \right) \\
   Z_6 & = & \rme^{\alpha\tau_p} \left( Z_{13} \sin(\gamma\tau_p) + Z_{14}\cos(\gamma\tau_p) \right) \\
Z_7 & = & \rme^{\alpha\tau_p} \left( Z_{13} \cos(\gamma\tau_p) - Z_{14}\sin(\gamma\tau_p) \right) - D_1 
+ \frac{4}{3}\frac{(K_1-2Q_1\alpha)}{(\alpha^2+\gamma^2)^{1/4}}\sin\left( \frac{\phi}{2} \right) \\
 & \, & +  \frac{4}{3}\frac{\left[-K_1\alpha + Q_1(\alpha^2-\gamma^2)\right]}{\gamma(\alpha^2+\gamma^2)^{1/4}}\cos\left( \frac{\phi}{2} \right)\\
Z_8 & = & \rme^{\alpha\tau_p} \left( Z_{13} \sin(\gamma\tau_p) + Z_{14}\cos(\gamma\tau_p) \right) - \frac{\left( E_1-D_1\alpha \right)}{\gamma} 
- \frac{4}{3}\frac{(K_1-2Q_1\alpha)}{(\alpha^2+\gamma^2)^{1/4}}\cos\left( \frac{\phi}{2} \right) \\
 & \, & +  \frac{4}{3}\frac{\left[-K_1\alpha + Q_1(\alpha^2-\gamma^2)\right]}{\gamma(\alpha^2+\gamma^2)^{1/4}}\sin\left( \frac{\phi}{2} \right)\\
  \end{eqnarray*}
where constants $Z_{11}$, $Z_{12}$, $Z_{13}$ and $Z_{14}$ read
 \begin{eqnarray*}
   Z_{11} & = & \cos(\Sto\tau_p)\left[ A_2 + \frac{4}{3}\sqrt{\frac{\Sto}{2}}F_2 + \frac{4}{3}\frac{G_2}{\sqrt{2\Sto}} \right] 
 - \Sto\sin(\Sto\tau_p) \left[ A_3 + \frac{4}{3}\sqrt{\frac{\Sto}{2}}F_3 + \frac{4}{3}\frac{G_3}{\sqrt{2\Sto}} \right] \\
 Z_{12} & = & \cos(\Sto\tau_p)\left[ \frac{B_2}{\Sto} - \frac{4}{3}\sqrt{\frac{\Sto}{2}}F_2 + \frac{4}{3}\frac{G_2}{\sqrt{2\Sto}} \right] 
  - \Sto\sin(\Sto\tau_p) \left[ \frac{B_3}{\Sto} - \frac{4}{3}\sqrt{\frac{\Sto}{2}}F_3 + \frac{4}{3}\frac{G_3}{\sqrt{2\Sto}} \right] \\
 Z_{13} & = & \cos(\Sto\tau_p)\left[ D_2 - \frac{4}{3}\frac{(K_2-2Q_2\alpha)}{(\alpha^2+\gamma^2)^{1/4}}\sin\left( \frac{\phi}{2} \right)
 - \frac{4}{3}\frac{\left[-K_2\alpha + Q_2(\alpha^2-\gamma^2)\right]}{\gamma(\alpha^2+\gamma^2)^{1/4}}\cos\left( \frac{\phi}{2} \right) \right] \\
  & \, &  - \Sto\sin(\Sto\tau_p) \left[ D_3 - \frac{4}{3}\frac{(K_3-2Q_3\alpha)}{(\alpha^2+\gamma^2)^{1/4}}\sin\left( \frac{\phi}{2} \right)
   - \frac{4}{3}\frac{\left[-K_3\alpha + Q_3(\alpha^2-\gamma^2)\right]}{\gamma(\alpha^2+\gamma^2)^{1/4}}\cos\left( \frac{\phi}{2} \right) \right] \\
 Z_{14} & = & \cos(\Sto\tau_p)\left[ \frac{\left( E_2-D_2\alpha \right)}{\gamma}  + \frac{4}{3}\frac{(K_2-2Q_2\alpha)}{(\alpha^2+\gamma^2)^{1/4}}\cos\left( \frac{\phi}{2} \right)
  - \frac{4}{3}\frac{\left[-K_2\alpha + Q_2(\alpha^2-\gamma^2)\right]}{\gamma(\alpha^2+\gamma^2)^{1/4}}\sin\left( \frac{\phi}{2} \right) \right] \\
   & \, &  - \Sto\sin(\Sto\tau_p) \left[ \frac{\left( E_3-D_3\alpha \right)}{\gamma}  + \frac{4}{3}\frac{(K_3-2Q_3\alpha)}{(\alpha^2+\gamma^2)^{1/4}}\cos\left( \frac{\phi}{2} \right)
     - \frac{4}{3}\frac{\left[-K_3\alpha + Q_3(\alpha^2-\gamma^2)\right]}{\gamma(\alpha^2+\gamma^2)^{1/4}}\sin\left( \frac{\phi}{2} \right) \right] 
\end{eqnarray*}
 The value for the former constants $S^s_{2,1,1}$ and $ S^{ns}_{2,1,1}$ is 
 \begin{eqnarray*}
 S^s_{2,1,1} & = & \frac{1}{2}b_r A_1 -\frac{1}{2}a_r B_1 + \frac{4}{3}F_1 \sqrt{\frac{\Sto}{2}}
 \left( \frac{1}{2}a_r \Sto + \frac{b_r}{2} \right) + \frac{4G_1}{3\sqrt{2\Sto}} 
 \left( -\frac{1}{2}a_r \Sto + \frac{b_r}{2} \right) \\
 S^{ns}_{2,1,1} & = & \frac{-\Sto}{2\alpha_r}\left( Z_1 b_r -Z_2 a_r\Sto \right) \\
  & \, &  + \frac{\Sto}{2\alpha}
 \left[ Z_3\sin(\Sto\tau_p) - Z_4\cos(\Sto\tau_p) \right]
 \left[ e_r\sin(\Sto\tau_p) + a_r\Sto\cos(\Sto\tau_p) \right] \\
 & \, & + \frac{\Sto}{2\alpha}
 \left[ Z_4\sin(\Sto\tau_p) + Z_3\cos(\Sto\tau_p) \right]
 \left[ \frac{f_r}{\Sto} \sin(\Sto\tau_p) + b_r\cos(\Sto\tau_p) \right]
 \end{eqnarray*}
\subsection{Constants for $\Wt^{(2)s}_{2,1,2}$ and $\Wt^{(2)ns}_{2,1,2}$ }
 \begin{eqnarray*}
 S^s_{2,1,2} & = & \frac{1}{2}b_r A_1 +\frac{1}{2}a_r B_1 + \frac{4}{3}F_1 \sqrt{\frac{\Sto}{2}}
 \left( \frac{b_r}{2} - \frac{1}{2}a_r \Sto \right) + \frac{4G_1}{3\sqrt{2\Sto}} 
 \left( \frac{1}{2}a_r \Sto + \frac{b_r}{2} \right) \\
 S^{ns}_{2,1,2} & = & \frac{-\Sto}{2\alpha_r} \left[ \cos(2\Sto\tau_p) \left( Z_2 a_r\Sto + Z_1 b_r \right) 
  + \sin(2\Sto\tau_p) \left( Z_2 b_r - Z_1 a_r\Sto \right) \right]\\
  & \, &  +\frac{\Sto}{2\alpha}\left[ e_r\sin(\Sto\tau_p) + a_r\Sto\cos(\Sto\tau_p) \right]\cdot \\
 & \, & \left\lbrace -\cos(\Sto\tau_p) \left[ Z_3\sin(2\Sto\tau_p) -Z_4\cos(2\Sto\tau_p) \right] 
 + \sin(\Sto\tau_p) \left[ Z_4\sin(2\Sto\tau_p) +Z_3\cos(2\Sto\tau_p) \right] \right\rbrace \\
 & \, & + \frac{\Sto}{2\alpha}\left[ \frac{f_r}{\Sto}\sin(\Sto\tau_p) + b_r\cos(\Sto\tau_p) \right]\cdot \\
 & \, & \left\lbrace \sin(\Sto\tau_p) \left[ Z_3\sin(2\Sto\tau_p) -Z_4\cos(2\Sto\tau_p) \right] 
 + \cos(\Sto\tau_p) \left[ Z_4\sin(2\Sto\tau_p) +Z_3\cos(2\Sto\tau_p) \right] \right\rbrace 
 \end{eqnarray*}
\subsection{Constants for $\Wt^{(2)s}_{2,1,3}$ and $\Wt^{(2)ns}_{2,1,3}$ }
 \begin{eqnarray*}
 S^s_{2,1,3} & = & \Sto \left[ b_r B_1 - a_r \Sto A_1 - \frac{4}{3}F_1 \sqrt{\frac{\Sto}{2}}
 \left( a_r \Sto + b_r \right) + \frac{4G_1}{3\sqrt{2\Sto}} 
 \left( -a_r \Sto + b_r\right) \right] \\
 S^{ns}_{2,1,3} & = & \frac{-\Sto^2}{\alpha_r}\left[\left( -Z_1 a_r\Sto + Z_2 b_r \right)\cos(2\Sto\tau_p)
  - \left( Z_1 b_r + Z_2 a_r\Sto \right)\sin(2\Sto\tau_p) \right] \\
 & \, & +\frac{\Sto^2}{\alpha}\left[ e_r\sin(\Sto\tau_p) + a_r\Sto\cos(\Sto\tau_p) \right]\cdot \\
 & \, & \left\lbrace -\cos(\Sto\tau_p) \left[ Z_4\sin(2\Sto\tau_p) +Z_3\cos(2\Sto\tau_p) \right] 
 + \sin(\Sto\tau_p) \left[ -Z_3\sin(2\Sto\tau_p) +Z_4\cos(2\Sto\tau_p) \right] \right\rbrace \\
 & \, & + \frac{\Sto^2}{\alpha}\left[ \frac{f_r}{\Sto}\sin(\Sto\tau_p) + b_r\cos(\Sto\tau_p) \right]\cdot \\
 & \, & \left\lbrace \sin(\Sto\tau_p) \left[ Z_4\sin(2\Sto\tau_p) +Z_3\cos(2\Sto\tau_p) \right] 
 + \cos(\Sto\tau_p) \left[ -Z_3\sin(2\Sto\tau_p) +Z_4\cos(2\Sto\tau_p) \right] \right\rbrace 
 \end{eqnarray*}
\subsection{Constants for $\Wt^{(2)s}_{2,1,4}$ and $\Wt^{(2)ns}_{2,1,4}$ }\label{T1_2}
Hereafter constants $T_1$ and $T_2$ will be used to refer to the following expressions:
\begin{eqnarray*}
T_1 & = & \frac{1}{\gamma}\left[ \left( E_2 - D_2\alpha \right)\cos(\Sto\tau_p) - \Sto\left( E_3 - D_3\alpha \right)\sin(\Sto\tau_p) \right] \\
& \, & +\frac{4}{3}\frac{\cos(\phi /2)}{\left( \alpha^2 + \gamma^2 \right)^{1/4}}\left[ \left( K_2 -2\alpha Q_2 \right)\cos(\Sto\tau_p)
 - \Sto\left( K_3 - 2\alpha Q_3 \right)\sin(\Sto\tau_p) \right] \\
& \, & +\frac{4}{3}\frac{\sin(\phi /2)}{\gamma\left( \alpha^2 + \gamma^2 \right)^{1/4}}
\left[ \left( \alpha K_2 -(\alpha^2-\gamma^2) Q_2 \right)\cos(\Sto\tau_p) - \Sto\left( \alpha K_3 -(\alpha^2-\gamma^2) Q_3 \right)\sin(\Sto\tau_p) \right]\\
T_2 & = & D_2\cos(\Sto\tau_p) -\Sto D_3 \sin(\Sto\tau_p) \\
& \, & -\frac{4}{3}\frac{\sin(\phi /2)}{\left( \alpha^2 + \gamma^2 \right)^{1/4}}
\left[ \left( K_2 -2\alpha Q_2 \right) \cos(\Sto\tau_p) - \Sto \left( K_3 -2\alpha Q_3 \right)\sin(\Sto\tau_p) \right] \\
& \, & +\frac{4}{3}\frac{\cos(\phi /2)}{\gamma\left( \alpha^2 + \gamma^2 \right)^{1/4}}
\left[  \left( \alpha K_2 -(\alpha^2-\gamma^2) Q_2 \right) \cos(\Sto\tau_p) - \Sto\left( \alpha K_3 -(\alpha^2-\gamma^2) Q_3 \right)\sin(\Sto\tau_p) \right] \\
\end{eqnarray*}
where $\phi= \tan^{-1}(-\frac{\gamma}{\alpha})$.
 \begin{eqnarray*}
 S^s_{2,1,4} & = & \left( \frac{d_r-2c_r\delta}{2}\right) A_1 
 - \left( \frac{c_r\left( \delta^2-\Omega^2 \right) -\delta d_r}{2\Omega}\right)
 \frac{B_1}{\Sto} \\
 & \, & + \frac{4}{3}F_1 \sqrt{\frac{\Sto}{2}}
 \left[ \left( \frac{d_r-2c_r\delta}{2}\right) + \left( \frac{c_r\left( \delta^2-\Omega^2 \right) -\delta d_r}{2\Omega}\right) \right] \\
 & \, & + \frac{4G_1}{3\sqrt{2\Sto}} 
  \left[ \left( \frac{d_r-2c_r\delta}{2}\right) - \left( \frac{c_r\left( \delta^2-\Omega^2 \right) -\delta d_r}{2\Omega}\right) \right]  \\
 S^{ns}_{2,1,4} & = & -\rme^{-\delta\tau_p}\frac{\Sto}{\alpha_r}\cos\left( (\Omega+\Sto)\tau_p\right)
 \left[ Z_1 \left( \frac{d_r-2c_r\delta}{2}\right) -  Z_2\left( \frac{c_r\left( \delta^2-\Omega^2 \right) -\delta d_r}{2\Omega}\right) \right] \\
  & \, &  -\rme^{-\delta\tau_p}\frac{\Sto}{\alpha_r}\sin\left( (\Omega+\Sto)\tau_p\right) 
 \left[ Z_2 \left( \frac{d_r-2c_r\delta}{2}\right) + Z_1 \left( \frac{c_r\left( \delta^2-\Omega^2 \right) -\delta d_r}{2\Omega}\right) \right] \\
 & \, & +\frac{1}{2}\left[ Z_4\sin\left( (\Omega+\Sto)\tau_p\right) + Z_3\cos\left( (\Omega+\Sto)\tau_p\right) \right] \cdot \\
 & \, & \left[ T_1 \left( \Omega\cos(\Omega\tau_p) + \delta\sin(\Omega\tau_p) \right)
 + T_2 \left( -\delta\cos(\Omega\tau_p) + \Omega\sin(\Omega\tau_p) \right) \right] \\
 & \, & +\frac{1}{2}\left[ Z_3\sin\left( (\Omega+\Sto)\tau_p\right) - Z_4\cos\left( (\Omega+\Sto)\tau_p\right) \right] \cdot \\
 & \, & \left[ T_1 \left( \Omega\sin(\Omega\tau_p) - \delta\cos(\Omega\tau_p) \right)
 + T_2 \left( -\delta\sin(\Omega\tau_p) - \Omega\cos(\Omega\tau_p) \right) \right] \\
  \end{eqnarray*}
\subsection{Constants for $\Wt^{(2)s}_{2,1,5}$ and $\Wt^{(2)ns}_{2,1,5}$ }
\begin{eqnarray*}
  S^s_{2,1,5} & = & (\Omega+\Sto) \left[ \left( \frac{d_r-2c_r\delta}{2}\right) \frac{B_1}{\Sto} 
  + \left( \frac{c_r\left( \delta^2-\Omega^2 \right) -\delta d_r}{2\Omega}\right)
  A_1 \right]\\
  & \, & + \frac{4}{3}F_1 \sqrt{\frac{\Sto}{2}} (\Omega+\Sto)
  \left[ -\left( \frac{d_r-2c_r\delta}{2}\right) + \left( \frac{c_r\left( \delta^2-\Omega^2 \right) -\delta d_r}{2\Omega}\right) \right] \\
  & \, & + \frac{4G_1}{3\sqrt{2\Sto}} (\Omega+\Sto)
   \left[ \left( \frac{d_r-2c_r\delta}{2}\right) + \left( \frac{c_r\left( \delta^2-\Omega^2 \right) -\delta d_r}{2\Omega}\right) \right]  \\
  S^{ns}_{2,1,5} & = & -\rme^{-\delta\tau_p}\frac{\Sto}{\alpha_r}(\Omega+\Sto)\cos\left( (\Omega+\Sto)\tau_p\right)
  \left[ Z_2 \left( \frac{d_r-2c_r\delta}{2}\right) +  Z_1\left( \frac{c_r\left( \delta^2-\Omega^2 \right) -\delta d_r}{2\Omega}\right) \right] \\
   & \, & - \rme^{-\delta\tau_p}\frac{\Sto}{\alpha_r}(\Omega+\Sto)\sin\left( (\Omega+\Sto)\tau_p\right) 
  \left[ -Z_1\left( \frac{d_r-2c_r\delta}{2}\right) +  Z_2\left( \frac{c_r\left( \delta^2-\Omega^2 \right) -\delta d_r}{2\Omega}\right) \right] \\
  & \, & +\frac{(\Omega+\Sto)}{2}\left[ -Z_3\sin\left( (\Omega+\Sto)\tau_p\right) + Z_4\cos\left( (\Omega+\Sto)\tau_p\right) \right] \cdot \\
  & \, & \left[ T_1 \left( \Omega\cos(\Omega\tau_p) + \delta\sin(\Omega\tau_p) \right)
  + T_2 \left( -\delta\cos(\Omega\tau_p) + \Omega\sin(\Omega\tau_p) \right) \right] \\
  & \, & +\frac{(\Omega+\Sto)}{2}\left[ Z_4\sin\left( (\Omega+\Sto)\tau_p\right) + Z_3\cos\left( (\Omega+\Sto)\tau_p\right) \right] \cdot \\
  & \, & \left[ T_1 \left( \Omega\sin(\Omega\tau_p) - \delta\cos(\Omega\tau_p) \right)
  + T_2 \left( -\delta\sin(\Omega\tau_p) - \Omega\cos(\Omega\tau_p) \right) \right] \\
\end{eqnarray*}
\subsection{Constants for $\Wt^{(2)s}_{2,1,6}$ and $\Wt^{(2)ns}_{2,1,6}$ }
 \begin{eqnarray*}
 S^s_{2,1,6} & = & \left( \frac{d_r-2c_r\delta}{2}\right) A_1 
 + \left( \frac{c_r\left( \delta^2-\Omega^2 \right) -\delta d_r}{2\Omega}\right)
 \frac{B_1}{\Sto} \\
 & \, & + \frac{4}{3}F_1 \sqrt{\frac{\Sto}{2}}
 \left[ \left( \frac{d_r-2c_r\delta}{2}\right) - \left( \frac{c_r\left( \delta^2-\Omega^2 \right) -\delta d_r}{2\Omega}\right) \right] \\
 & \, & + \frac{4G_1}{3\sqrt{2\Sto}} 
  \left[ \left( \frac{d_r-2c_r\delta}{2}\right) + \left( \frac{c_r\left( \delta^2-\Omega^2 \right) -\delta d_r}{2\Omega}\right) \right]  \\
 S^{ns}_{2,1,6} & = & -\rme^{-\delta\tau_p}\frac{\Sto}{\alpha_r}\cos\left( (\Omega-\Sto)\tau_p\right)
 \left[ Z_1\left( \frac{d_r-2c_r\delta}{2}\right) +  Z_2\left( \frac{c_r\left( \delta^2-\Omega^2 \right) -\delta d_r}{2\Omega}\right) \right] \\
  & \, & - \rme^{-\delta\tau_p}\frac{\Sto}{\alpha_r}\sin\left( (\Omega-\Sto)\tau_p\right) 
 \left[ -Z_2\left( \frac{d_r-2c_r\delta}{2}\right) +  Z_1\left( \frac{c_r\left( \delta^2-\Omega^2 \right) -\delta d_r}{2\Omega}\right) \right] \\
 & \, & +\frac{1}{2}\left[ Z_3\cos\left( (\Omega-\Sto)\tau_p\right) - Z_4\sin\left( (\Omega+\Sto)\tau_p\right) \right] \cdot \\
 & \, & \left[ T_1 \left( \Omega\cos(\Omega\tau_p) + \delta\sin(\Omega\tau_p) \right)
 + T_2 \left( -\delta\cos(\Omega\tau_p) + \Omega\sin(\Omega\tau_p) \right) \right] \\
 & \, & +\frac{1}{2}\left[ Z_4\cos\left( (\Omega-\Sto)\tau_p\right) + Z_3\sin\left( (\Omega-\Sto)\tau_p\right) \right] \cdot \\
 & \, & \left[ T_1 \left( \Omega\sin(\Omega\tau_p) - \delta\cos(\Omega\tau_p) \right)
 + T_2 \left( -\delta\sin(\Omega\tau_p) - \Omega\cos(\Omega\tau_p) \right) \right] \\
  \end{eqnarray*}
\subsection{Constants for $\Wt^{(2)s}_{2,1,7}$ and $\Wt^{(2)ns}_{2,1,7}$ }
\begin{eqnarray*}
  S^s_{2,1,7} & = & (\Omega-\Sto) \left[ -\left( \frac{d_r-2c_r\delta}{2}\right) \frac{B_1}{\Sto} 
  + \left( \frac{c_r\left( \delta^2-\Omega^2 \right) -\delta d_r}{2\Omega}\right)
  A_1 \right]\\
  & \, & + \frac{4}{3}F_1 \sqrt{\frac{\Sto}{2}} (\Omega-\Sto)
  \left[ \left( \frac{d_r-2c_r\delta}{2}\right) + \left( \frac{c_r\left( \delta^2-\Omega^2 \right) -\delta d_r}{2\Omega}\right) \right] \\
  & \, & + \frac{4G_1}{3\sqrt{2\Sto}} (\Omega-\Sto)
   \left[ -\left( \frac{d_r-2c_r\delta}{2}\right) + \left( \frac{c_r\left( \delta^2-\Omega^2 \right) -\delta d_r}{2\Omega}\right) \right]  \\
  S^{ns}_{2,1,7} & = & -\rme^{-\delta\tau_p}\frac{\Sto}{\alpha_r}(\Omega-\Sto)\cos\left( (\Omega-\Sto)\tau_p\right)
  \left[ -Z_2\left( \frac{d_r-2c_r\delta}{2}\right) +  Z_1\left( \frac{c_r\left( \delta^2-\Omega^2 \right) -\delta d_r}{2\Omega}\right) \right] \\
   & \, & - \rme^{-\delta\tau_p}\frac{\Sto}{\alpha_r}(\Omega-\Sto)\sin\left( (\Omega-\Sto)\tau_p\right) 
  \left[ -Z_1\left( \frac{d_r-2c_r\delta}{2}\right) - Z_2 \left( \frac{c_r\left( \delta^2-\Omega^2 \right) -\delta d_r}{2\Omega}\right) \right] \\
  & \, & +\frac{(\Omega-\Sto)}{2}\left[ -Z_3\sin\left( (\Omega-\Sto)\tau_p\right) -Z_4 \cos\left( (\Omega-\Sto)\tau_p\right) \right] \cdot \\
  & \, & \left[ T_1 \left( \Omega\cos(\Omega\tau_p) + \delta\sin(\Omega\tau_p) \right)
  + T_2 \left( -\delta\cos(\Omega\tau_p) + \Omega\sin(\Omega\tau_p) \right) \right] \\
  & \, & +\frac{(\Omega-\Sto)}{2}\left[ -Z_4\sin\left( (\Omega-\Sto)\tau_p\right) +Z_3 \cos\left( (\Omega-\Sto)\tau_p\right) \right] \cdot \\
  & \, & \left[ T_1 \left( \Omega\sin(\Omega\tau_p) - \delta\cos(\Omega\tau_p) \right)
  + T_2 \left( -\delta\sin(\Omega\tau_p) - \Omega\cos(\Omega\tau_p) \right) \right] \\
\end{eqnarray*}
\subsection{Constants for $\Wt^{(2)s}_{2,2,1}$ and $\Wt^{(2)ns}_{2,2,1}$ }
Constants involved in this second part of the first-order velocity do not coincide with the constants given in previous subsections. 
Therefore, here we will provide with the linear systems whose solution allows us to obtain constants $A_{2,2,k}$, $B_{2,2,k}$, $C_{2,2,k}$, ... for each component $k$.\\
In particular, the linear system associated to component $k=1$ reads,
 \begin{equation*}
  \left[\begin{array}{c}
  A_{2,2,1}\\
  B_{2,2,1}\\
  C_{2,2,1}\\
  D_{2,2,1}\\
  E_{2,2,1}
  \end{array}\right]
   = (M_{1,4}(\alpha,\gamma + \Sto))^{-1} \times
  \left[\begin{array}{c}
   0\\
   1\\
   1+\alpha\\
   -2/3+\alpha\\
   -2\alpha/3
    \end{array}\right]
   , \; 
  \left[\begin{array}{c}
   F_{2,2,1}\\
   G_{2,2,1}\\
   H_{2,2,1}\\
   Q_{2,2,1}\\
   K_{2,2,1}
    \end{array}\right]
    =(M_{1,4}(\alpha, \gamma + \Sto))^{-1} \times
  \left[\begin{array}{c}
     0\\
     0\\
     0\\
     1\\
     \alpha
      \end{array}\right]
   \end{equation*}
 where $\varphi_1=\alpha$ and $\varphi_2=\gamma + \Sto$ in matrix $M_{1,4}(\varphi_1, \varphi_2)$.\\
Constants $S^s_{2,2,k}$ and $S^{ns}_{2,2,k}$ are also different for each component. Their value for the former is 
 \begin{eqnarray*}
  S^s_{2,2,1} & = & \frac{b_r}{2} D_1 +\frac{a_r \Sto}{2}\frac{\left( E_1-D_1\alpha \right)}{\gamma} 
  + \frac{4}{3} \frac{\left( K_1 -2Q_1\alpha \right)}{\left( \alpha^2+\gamma^2\right)^{1/4}} 
  \left( -\frac{b_r}{2}\sin(\phi/2)+\frac{a_r\Sto}{2}\cos(\phi/2) \right)  \\
  & \, & + \frac{4}{3} \frac{\left( -K_1 \alpha +Q_1\left(\alpha^2-\gamma^2 \right) \right)}{\gamma \left( \alpha^2+\gamma^2\right)^{1/4}}
   \left( -\frac{b_r}{2}\cos(\phi/2)-\frac{a_r\Sto}{2}\sin(\phi/2) \right)  \\
  S^{ns}_{2,2,1} & = & \frac{-\Sto}{2\alpha_r}\rme^{-\alpha\tau_p} \left[ \left( Z_6 b_r - Z_5 a_r\Sto \right)
   \sin \left((\gamma + \Sto)\tau_p \right)+ \left( Z_5 b_r + Z_6 a_r\Sto \right) \cos \left((\gamma + \Sto)\tau_p \right) \right]\\
  & \, & + \frac{\rme^{-\alpha\tau_p}}{2} \left[ \left( T_3 Z_7+T_4 Z_8 \right)\sin \left((\gamma + \Sto)\tau_p \right) 
 + \left( T_4 Z_7 - T_3 Z_8 \right)\cos \left((\gamma + \Sto)\tau_p \right) 
  \right]  
  \end{eqnarray*}
  where hereafter constants $T_3$ and $T_4$ will refer to the following expressions
 \begin{eqnarray*}
  T_3 & = & -\frac{\Sto}{\alpha}\cos(\Sto\tau_p) \left[ e_r\sin(\Sto\tau_p) + a_r\Sto\cos(\Sto\tau_p) \right]  \\
  & \, & + \frac{\Sto}{\alpha}\sin(\Sto\tau_p) \left[ \frac{f_r}{\Sto}\sin(\Sto\tau_p) + b_r\cos(\Sto\tau_p) \right]   \\
  T_4 & = & \frac{\Sto}{\alpha}\sin(\Sto\tau_p) \left[ e_r\sin(\Sto\tau_p) + a_r\Sto\cos(\Sto\tau_p) \right]  \\
    & \, & + \frac{\Sto}{\alpha}\cos(\Sto\tau_p) \left[ \frac{f_r}{\Sto}\sin(\Sto\tau_p) + b_r\cos(\Sto\tau_p) \right]  
  \end{eqnarray*}
 \subsection{Constants for $\Wt^{(2)s}_{2,2,2}$ and $\Wt^{(2)ns}_{2,2,2}$ } 
  \begin{equation*}
    \left[\begin{array}{c}
    A_{2,2,2}\\
    B_{2,2,2}\\
    C_{2,2,2}\\
    D_{2,2,2}\\
    E_{2,2,2}
    \end{array}\right]
     = (M_{1,4}(\alpha, \gamma + \Sto))^{-1} \times
    \left[\begin{array}{c}
     0\\
     0\\
     1\\
     1\\
     -2/3
      \end{array}\right]
    , \;      
    \left[\begin{array}{c}
     F_{2,2,2}\\
     G_{2,2,2}\\
     H_{2,2,2}\\
     Q_{2,2,2}\\
     K_{2,2,2}
      \end{array}\right]
      =(M_{1,4}(\alpha, \gamma + \Sto))^{-1} \times
    \left[\begin{array}{c}
       0\\
       0\\
       0\\
       0\\
       1
        \end{array}\right]
     \end{equation*}
where $\varphi_1=\alpha$ and $\varphi_2=\gamma + \Sto$ in matrix $M_{1,4}(\varphi_1, \varphi_2)$.\\
Constants $S^s_{2,2,2}$ and $S^{ns}_{2,2,2}$ read
 \begin{eqnarray*}
  S^s_{2,2,2} & = & \left( \gamma+\Sto \right) \left[ -\frac{a_r \Sto}{2} D_1 +\frac{b_r}{2}\frac{\left( E_1-D_1\alpha \right)}{\gamma} 
  + \frac{4}{3} \frac{\left( K_1 -2Q_1\alpha \right)}{\left( \alpha^2+\gamma^2\right)^{1/4}} 
  \left( \frac{a_r\Sto}{2}\sin\left(\frac{\phi}{2}\right)+ \frac{b_r}{2}\cos\left(\frac{\phi}{2}\right) \right) \right] \\
  & \, & + \left( \gamma+\Sto \right)\frac{4}{3} \frac{\left( -K_1 \alpha +Q_1\left(\alpha^2-\gamma^2 \right) \right)}{\gamma \left( \alpha^2+\gamma^2\right)^{1/4}}
   \left(\frac{a_r\Sto}{2}\cos\left(\frac{\phi}{2}\right)-\frac{b_r}{2}\sin\left(\frac{\phi}{2}\right) \right)  \\
  S^{ns}_{2,2,2} & = & -\frac{\Sto}{\alpha_r}\frac{\left( \gamma+\Sto \right)}{2}\rme^{-\alpha\tau_p}
  \left( -Z_5 b_r -Z_6 a_r\Sto \right) \sin \left((\gamma + \Sto)\tau_p \right) \\
  & \, & -\frac{\Sto}{\alpha_r}\frac{\left( \gamma+\Sto \right)}{2}\rme^{-\alpha\tau_p}
    \left( Z_6 b_r -Z_5 a_r\Sto \right) \cos \left((\gamma + \Sto)\tau_p \right) \\
  & \, & + \frac{\left( \gamma+\Sto \right)}{2}\rme^{-\alpha\tau_p} \left[ \left( -T_4Z_7 + T_3Z_8\right) 
  \sin \left((\gamma + \Sto)\tau_p \right) + 
  \left( T_3Z_7 + T_4Z_8\right)\cos \left((\gamma + \Sto)\tau_p \right)\right]
  \end{eqnarray*}

  \subsection{Constants for $\Wt^{(2)s}_{2,2,3}$ and $\Wt^{(2)ns}_{2,2,3}$ } 
 \begin{equation*}
   \left[\begin{array}{c}
   A_{2,2,3}\\
   B_{2,2,3}\\
   C_{2,2,3}\\
   D_{2,2,3}\\
   E_{2,2,3}
   \end{array}\right]
    = (M_{1,4}(\alpha, \gamma-\Sto))^{-1} \times
   \left[\begin{array}{c}
    0\\
    1\\
    1+\alpha\\
    -2/3+\alpha\\
    -2\alpha/3
     \end{array}\right]
    , \; 
   \left[\begin{array}{c}
    F_{2,2,3}\\
    G_{2,2,3}\\
    H_{2,2,3}\\
    Q_{2,2,3}\\
    K_{2,2,3}
     \end{array}\right]
     =(M_{1,4}(\alpha, \gamma-\Sto))^{-1} \times
   \left[\begin{array}{c}
      0\\
      0\\
      0\\
      1\\
      \alpha
       \end{array}\right]
    \end{equation*}  
 where $\varphi_1=\alpha$ and $\varphi_2=\gamma - \Sto$ in matrix $M_{1,4}(\varphi_1, \varphi_2)$.\\
 Constants $S^s_{2,2,3}$ and $S^{ns}_{2,2,3}$ are defined as
 \begin{eqnarray*}
  S^s_{2,2,3} & = & \frac{b_r}{2} D_1 -\frac{a_r \Sto}{2}\frac{\left( E_1-D_1\alpha \right)}{\gamma} 
  + \frac{4}{3} \frac{\left( K_1 -2Q_1\alpha \right)}{\left( \alpha^2+\gamma^2\right)^{1/4}} 
  \left( -\frac{b_r}{2}\sin(\phi/2)-\frac{a_r\Sto}{2}\cos(\phi/2) \right)  \\
  & \, & + \frac{4}{3} \frac{\left( -K_1 \alpha +Q_1\left(\alpha^2-\gamma^2 \right) \right)}{\gamma \left( \alpha^2+\gamma^2\right)^{1/4}}
   \left( -\frac{b_r}{2}\cos(\phi/2)+\frac{a_r\Sto}{2}\sin(\phi/2) \right)  \\
  S^{ns}_{2,2,3} & = & \frac{-\Sto}{2\alpha_r}\rme^{-\alpha\tau_p}\left[ \left( Z_6 b_r + Z_5a_r\Sto \right) \sin \left((\gamma - \Sto)\tau_p \right)+ \left( Z_5 b_r - Z_6 a_r\Sto \right) \cos \left((\gamma - \Sto)\tau_p \right) \right]\\
   & \, & +\frac{\rme^{-\alpha\tau_p}}{2} 
   \left[ \left( -T_3 Z_7 +T_4 Z_8 \right)\sin\left((\gamma - \Sto)\tau_p \right) 
    + \left( T_4 Z_7 +T_3 Z_8 \right)\cos\left((\gamma - \Sto)\tau_p \right) \right]
  \end{eqnarray*}

  \subsection{Constants for $\Wt^{(2)s}_{2,2,4}$ and $\Wt^{(2)ns}_{2,2,4}$ } 
    \begin{equation*}
      \left[\begin{array}{c}
      A_{2,2,4}\\
      B_{2,2,4}\\
      C_{2,2,4}\\
      D_{2,2,4}\\
      E_{2,2,4}
      \end{array}\right]
       = (M_{1,4}(\alpha, \gamma - \Sto))^{-1} \times
      \left[\begin{array}{c}
       0\\
       0\\
       1\\
       1\\
       -2/3
        \end{array}\right]
      , \;      
      \left[\begin{array}{c}
       F_{2,2,4}\\
       G_{2,2,4}\\
       H_{2,2,4}\\
       Q_{2,2,4}\\
       K_{2,2,4}
        \end{array}\right]
        =(M_{1,4}(\alpha, \gamma - \Sto))^{-1} \times
      \left[\begin{array}{c}
         0\\
         0\\
         0\\
         0\\
         1
          \end{array}\right]
       \end{equation*}
  where $\varphi_1=\alpha$ and $\varphi_2=\gamma - \Sto$ in matrix $M_{1,4}(\varphi_1, \varphi_2)$.\\
  Constants $S^s_{2,2,4}$ and $S^{ns}_{2,2,4}$ are
   \begin{eqnarray*}
    S^s_{2,2,4} & = & \left( \gamma-\Sto \right) \left[ \frac{a_r \Sto}{2} D_1 +\frac{b_r}{2}\frac{\left( E_1-D_1\alpha \right)}{\gamma}\right] \\
    & \, & + \left( \gamma-\Sto \right) \frac{4}{3} \frac{\left( K_1 -2Q_1\alpha \right)}{\left( \alpha^2+\gamma^2\right)^{1/4}} 
    \left[- \frac{a_r\Sto}{2}\sin\left(\frac{\phi}{2}\right)+ \frac{b_r}{2}\cos\left(\frac{\phi}{2}\right) \right]  \\
    & \, & + \left( \gamma-\Sto \right)\frac{4}{3} \frac{\left( -K_1 \alpha +Q_1\left(\alpha^2-\gamma^2 \right) \right)}{\gamma \left( \alpha^2+\gamma^2\right)^{1/4}}
     \left[-\frac{a_r\Sto}{2}\cos\left(\frac{\phi}{2}\right)-\frac{b_r}{2}\sin\left(\frac{\phi}{2}\right) \right]  \\
    S^{ns}_{2,2,4} & = & -\frac{\Sto}{\alpha_r}\frac{\left( \gamma-\Sto \right)}{2}\rme^{-\alpha\tau_p}
     \left( -Z_5 b_r +Z_6 a_r\Sto \right) \sin \left((\gamma - \Sto)\tau_p \right)\\
& \, & - \frac{\Sto}{\alpha_r}\frac{\left( \gamma-\Sto \right)}{2}\rme^{-\alpha\tau_p} \left( Z_6 b_r + Z_5 a_r\Sto \right) \cos \left((\gamma - \Sto)\tau_p \right) \\
& \, & + \frac{\left( \gamma-\Sto \right)}{2}\rme^{-\alpha\tau_p} \left[
\left( -T_4 Z_7 - T_3 Z_8 \right)\sin \left((\gamma - \Sto)\tau_p \right) +
\left( -T_3 Z_7 + T_4 Z_8 \right)\cos \left((\gamma - \Sto)\tau_p \right) \right]
    \end{eqnarray*}
  \subsection{Constants for $\Wt^{(2)s}_{2,2,5}$ and $\Wt^{(2)ns}_{2,2,5}$ }    
 \begin{equation*}
  \left[\begin{array}{c}
  A_{2,2,5}\\
  B_{2,2,5}\\
  C_{2,2,5}\\
  D_{2,2,5}\\
  E_{2,2,5}
  \end{array}\right]
   = (M_{1,4}(\alpha+\delta,\Omega + \gamma))^{-1} \times
  \left[\begin{array}{c}
   0\\
   1\\
   1+\left( \alpha+\delta\right)\\
   -2/3+\left( \alpha+\delta\right)\\
   -2\left( \alpha+\delta\right)/3
    \end{array}\right]
   \end{equation*}
    \begin{equation*}
  \left[\begin{array}{c}
   F_{2,2,5}\\
   G_{2,2,5}\\
   H_{2,2,5}\\
   Q_{2,2,5}\\
   K_{2,2,5}
    \end{array}\right]
    =(M_{1,4}(\alpha+\delta,\Omega + \gamma))^{-1} \times
  \left[\begin{array}{c}
     0\\
     0\\
     0\\
     1\\
     \ \left( \alpha+\delta\right)
      \end{array}\right]
   \end{equation*}
 where $\varphi_1= \alpha+\delta$ and $\varphi_2=\Omega+\gamma $ in matrix $M_{1,4}(\varphi_1, \varphi_2)$.\\
Constants $S^s_{2,2,5}$ and $S^{ns}_{2,2,5}$ have the following value
 \begin{eqnarray*}
  S^s_{2,2,5} & = & \left( \frac{d_r-2c_r\delta}{2} \right) D_1 
  -\left( \frac{c_r( \delta ^2-\Omega ^2 )-\delta d_r}{2\Omega} \right) \frac{\left( E_1-D_1\alpha \right)}{\gamma}\\
  & \, & + \frac{4}{3} \frac{\left( K_1 -2Q_1\alpha \right)}{\left( \alpha^2+\gamma^2\right)^{1/4}} 
  \left[ -\left( \frac{d_r-2c_r\delta}{2} \right) \sin(\phi/2)
  -\left( \frac{c_r( \delta ^2-\Omega ^2 )-\delta d_r}{2\Omega} \right) \cos(\phi/2) \right]  \\
  & \, & + \frac{4}{3} \frac{\left( -K_1 \alpha +Q_1\left(\alpha^2-\gamma^2 \right) \right)}{\gamma \left( \alpha^2+\gamma^2\right)^{1/4}}
   \left[ - \left( \frac{d_r-2c_r\delta}{2} \right)\cos(\phi/2)
    +\left( \frac{c_r( \delta ^2-\Omega ^2 )-\delta d_r}{2\Omega} \right) \sin(\phi/2) \right]  \\
  S^{ns}_{2,2,5} & = & \frac{-\Sto}{\alpha_r}\rme^{-(\alpha+\delta)\tau_p}
  \left[ Z_5\left( \frac{d_r-2c_r\delta}{2} \right)
  - Z_6\left( \frac{c_r( \delta ^2-\Omega ^2 )-\delta d_r}{2\Omega} \right)\right]
  \cos\left((\Omega+\gamma)\tau_p\right)  \\
  & \, & + \frac{-\Sto}{\alpha_r} \rme^{-(\alpha+\delta)\tau_p}\left[ Z_6\left( \frac{d_r-2c_r\delta}{2} \right)
  + Z_5\left( \frac{c_r( \delta ^2-\Omega ^2 )-\delta d_r}{2\Omega} \right)\right]
    \sin\left((\Omega+\gamma)\tau_p\right)  \\
  & \, & + \frac{ \rme^{-\alpha\tau_p}}{2}
  \left[ Z_7\cos\left((\Omega+\gamma)\tau_p\right) + Z_8\sin\left((\Omega+\gamma)\tau_p\right) \right]\cdot \\
  & \, & \left[ T_1 \left( \Omega\cos(\Omega\tau_p) + \delta\sin(\Omega\tau_p) \right)
    + T_2 \left( -\delta\cos(\Omega\tau_p) + \Omega\sin(\Omega\tau_p) \right) \right] \\
& \, & + \frac{ \rme^{-\alpha\tau_p}}{2}
\left[ -Z_8\cos\left((\Omega+\gamma)\tau_p\right) + Z_7\sin\left((\Omega+\gamma)\tau_p\right) \right]\cdot \\
& \, & \left[ T_1 \left( \Omega\sin(\Omega\tau_p) - \delta\cos(\Omega\tau_p) \right) 
+ T_2 \left( -\delta\sin(\Omega\tau_p) - \Omega\cos(\Omega\tau_p) \right) \right] .
  \end{eqnarray*}
Note that constants $T_1$ and $T_2$ have been defined in subsection \ref{T1_2}.

  \subsection{Constants for $\Wt^{(2)s}_{2,2,6}$ and $\Wt^{(2)ns}_{2,2,6}$ }    
  \begin{equation*}
      \left[\begin{array}{c}
      A_{2,2,6}\\
      B_{2,2,6}\\
      C_{2,2,6}\\
      D_{2,2,6}\\
      E_{2,2,6}
      \end{array}\right]
       = (M_{1,4}(\alpha+\delta, \Omega+\gamma))^{-1} \times
      \left[\begin{array}{c}
       0\\
       0\\
       1\\
       1\\
       -2/3
        \end{array}\right]
      , \;      
      \left[\begin{array}{c}
       F_{2,2,6}\\
       G_{2,2,6}\\
       H_{2,2,6}\\
       Q_{2,2,6}\\
       K_{2,2,6}
        \end{array}\right]
        =(M_{1,4}(\alpha+\delta, \Omega+\gamma))^{-1} \times
      \left[\begin{array}{c}
         0\\
         0\\
         0\\
         0\\
         1
          \end{array}\right]
       \end{equation*}
   where $\varphi_1= \alpha+\delta$ and $\varphi_2=\Omega+\gamma $ in matrix $M_{1,4}(\varphi_1, \varphi_2)$.\\
  Constants $S^s_{2,2,6}$ and $S^{ns}_{2,2,6}$ are defined as
   \begin{eqnarray*}
    S^s_{2,2,6} & = & (\Omega+\gamma) \left[ \left( \frac{d_r-2c_r\delta}{2} \right) \frac{\left( E_1-D_1\alpha \right)}{\gamma}   
    +\left( \frac{c_r( \delta ^2-\Omega ^2 )-\delta d_r}{2\Omega} \right) D_1 \right]\\
    & \, & + (\Omega+\gamma) \frac{4}{3} \frac{\left( K_1 -2Q_1\alpha \right)}{\left( \alpha^2+\gamma^2\right)^{1/4}} 
    \left[ \left( \frac{d_r-2c_r\delta}{2} \right) \cos(\phi/2)
    -\left( \frac{c_r( \delta ^2-\Omega ^2 )-\delta d_r}{2\Omega} \right) \sin(\phi/2) \right]  \\
    & \, & - (\Omega+\gamma)\frac{4}{3} \left[ \frac{ -K_1 \alpha +Q_1\left(\alpha^2-\gamma^2 \right) }{\gamma \left( \alpha^2+\gamma^2\right)^{1/4}}\right]
    \left( \frac{d_r-2c_r\delta}{2} \right)\sin(\phi/2) \\
     & \, &  - (\Omega+\gamma)\frac{4}{3} \left[ \frac{ -K_1 \alpha +Q_1\left(\alpha^2-\gamma^2 \right) }{\gamma \left( \alpha^2+\gamma^2\right)^{1/4}}\right]
   \left( \frac{c_r( \delta ^2-\Omega ^2 )-\delta d_r}{2\Omega} \right) \cos(\phi/2) \\
    S^{ns}_{2,2,6} & = & -\frac{\Sto}{\alpha_r}(\Omega+\gamma)\rme^{-(\alpha+\delta)\tau_p}
    \left[ Z_6\left( \frac{d_r-2c_r\delta}{2} \right) 
    + Z_5\left( \frac{c_r( \delta ^2-\Omega ^2 )-\delta d_r}{2\Omega}\right)  \right]
    \cos\left((\Omega+\gamma)\tau_p\right)  \\
    & \, & -\frac{\Sto}{\alpha_r} (\Omega+\gamma)\rme^{-(\alpha+\delta)\tau_p}
    \left[  -Z_5\left( \frac{d_r-2c_r\delta}{2} \right) 
    + Z_6\left( \frac{c_r( \delta ^2-\Omega ^2 )-\delta d_r}{2\Omega}\right) \right] 
      \sin\left((\Omega+\gamma)\tau_p\right)\\
    & \, & + \frac{(\Omega+\gamma)}{2}\rme^{-\alpha\tau_p}
    \left[Z_8 \cos\left((\Omega+\gamma)\tau_p\right) - Z_7\sin\left((\Omega+\gamma)\tau_p\right) \right]\cdot \\
    & \, & \left[ T_1 \left( \Omega\cos(\Omega\tau_p) + \delta\sin(\Omega\tau_p) \right)
      + T_2 \left( -\delta\cos(\Omega\tau_p) + \Omega\sin(\Omega\tau_p) \right) \right] \\
  & \, & + \frac{(\Omega+\gamma)}{2}\rme^{-\alpha\tau_p}
  \left[ Z_7\cos\left((\Omega+\gamma)\tau_p\right) +Z_8 \sin\left((\Omega+\gamma)\tau_p\right) \right]\cdot \\
  & \, & \left[ T_1 \left( \Omega\sin(\Omega\tau_p) - \delta\cos(\Omega\tau_p) \right) 
  + T_2 \left( -\delta\sin(\Omega\tau_p) - \Omega\cos(\Omega\tau_p) \right) \right] 
    \end{eqnarray*}

  \subsection{Constants for $\Wt^{(2)s}_{2,2,7}$ and $\Wt^{(2)ns}_{2,2,7}$ }    
 \begin{equation*}
    \left[\begin{array}{c}
    A_{2,2,7}\\
    B_{2,2,7}\\
    C_{2,2,7}\\
    D_{2,2,7}\\
    E_{2,2,7}
    \end{array}\right]
     = (M_{1,4}(\alpha+\delta, \Omega- \gamma))^{-1} \times
    \left[\begin{array}{c}
     0\\
     1\\
     1+(\alpha+\delta)\\
     -2/3+(\alpha+\delta)\\
     -2(\alpha+\delta)/3
      \end{array}\right],
     \end{equation*}  
 \begin{equation*}     
    \left[\begin{array}{c}
     F_{2,2,7}\\
     G_{2,2,7}\\
     H_{2,2,7}\\
     Q_{2,2,7}\\
     K_{2,2,7}
      \end{array}\right]
      =(M_{1,4}(\alpha+\delta, \Omega- \gamma))^{-1} \times
    \left[\begin{array}{c}
       0\\
       0\\
       0\\
       1\\
       (\alpha+\delta)
        \end{array}\right]
     \end{equation*}  
  where $\varphi_1=\alpha+\delta$ and $\varphi_2=\Omega -\gamma$ in matrix $M_{1,4}(\varphi_1, \varphi_2)$.\\  
      Constants $S^s_{2,2,7}$ and $S^{ns}_{2,2,7}$ are given by the expressions below
   \begin{eqnarray*}
       S^s_{2,2,7} & = & \left( \frac{d_r-2c_r\delta}{2} \right) D_1 
       +\left( \frac{c_r( \delta ^2-\Omega ^2 )-\delta d_r}{2\Omega} \right) \frac{\left( E_1-D_1\alpha \right)}{\gamma}\\
       & \, & + \frac{4}{3} \frac{\left( K_1 -2Q_1\alpha \right)}{\left( \alpha^2+\gamma^2\right)^{1/4}} 
       \left[ -\left( \frac{d_r-2c_r\delta}{2} \right) \sin(\phi/2)
       +\left( \frac{c_r( \delta ^2-\Omega ^2 )-\delta d_r}{2\Omega} \right) \cos(\phi/2) \right]  \\
       & \, & - \frac{4}{3} \left[\frac{\left( -K_1 \alpha +Q_1\left(\alpha^2-\gamma^2 \right) \right)}{\gamma \left( \alpha^2+\gamma^2\right)^{1/4}} \right]
        \left( \frac{d_r-2c_r\delta}{2} \right)\cos(\phi/2) \\
       & \, & - \frac{4}{3} \left[\frac{\left( -K_1 \alpha +Q_1\left(\alpha^2-\gamma^2 \right) \right)}{\gamma \left( \alpha^2+\gamma^2\right)^{1/4}} \right]        
         \left( \frac{c_r( \delta ^2-\Omega ^2 )-\delta d_r}{2\Omega} \right) \sin(\phi/2) \\
  S^{ns}_{2,2,7} & = &-\frac{\Sto}{\alpha_r} \rme^{-(\alpha+\delta)\tau_p}
  \left[ Z_5 \left( \frac{d_r-2c_r\delta}{2}\right) + Z_6\left(\frac{c_r( \delta ^2-\Omega ^2 )-\delta d_r}{2\Omega}\right) \right]
    \cos\left((\Omega-\gamma)\tau_p\right) \\
    & \, & - \frac{\Sto}{\alpha_r}\rme^{-(\alpha+\delta)\tau_p}
    \left[ -Z_6\left( \frac{d_r-2c_r\delta}{2}\right) +Z_5 \left(\frac{c_r( \delta ^2-\Omega ^2 )-\delta d_r}{2\Omega}\right) \right]
      \sin\left((\Omega-\gamma)\tau_p\right)  \\
    & \, & + \frac{ \rme^{-\alpha\tau_p}}{2}\left[ Z_7\cos\left((\Omega-\gamma)\tau_p\right) -Z_8\sin\left((\Omega-\gamma)\tau_p\right) \right]\cdot \\
    & \, & \left[ T_1 \left( \Omega\cos(\Omega\tau_p) + \delta\sin(\Omega\tau_p) \right)
      + T_2 \left( -\delta\cos(\Omega\tau_p) + \Omega\sin(\Omega\tau_p) \right) \right] \\
  & \, & + \frac{ \rme^{-\alpha\tau_p}}{2}\left[ Z_8\cos\left((\Omega-\gamma)\tau_p\right) 
  + Z_7\sin\left((\Omega-\gamma)\tau_p\right) \right]\cdot \\
  & \, & \left[ T_1 \left( \Omega\sin(\Omega\tau_p) - \delta\cos(\Omega\tau_p) \right) 
  + T_2 \left( -\delta\sin(\Omega\tau_p) - \Omega\cos(\Omega\tau_p) \right) \right] .
    \end{eqnarray*} 

  \subsection{Constants for $\Wt^{(2)s}_{2,2,8}$ and $\Wt^{(2)ns}_{2,2,8}$ } 
      \begin{equation*}
        \left[\begin{array}{c}
        A_{2,2,8}\\
        B_{2,2,8}\\
        C_{2,2,8}\\
        D_{2,2,8}\\
        E_{2,2,8}
        \end{array}\right]
         = (M_{1,4}(\alpha+\delta, \Omega-\gamma))^{-1} \times
        \left[\begin{array}{c}
         0\\
         0\\
         1\\
         1\\
         -2/3
          \end{array}\right]
        , \;      
        \left[\begin{array}{c}
         F_{2,2,8}\\
         G_{2,2,8}\\
         H_{2,2,8}\\
         Q_{2,2,8}\\
         K_{2,2,8}
          \end{array}\right]
          =(M_{1,4}(\alpha+\delta, \Omega-\gamma))^{-1} \times
        \left[\begin{array}{c}
           0\\
           0\\
           0\\
           0\\
           1
            \end{array}\right]
         \end{equation*}
 where $\varphi_1=\alpha+\delta$ and $\varphi_2=\Omega -\gamma$ in matrix $M_{1,4}(\varphi_1, \varphi_2)$.\\  
Constants $S^s_{2,2,8}$ and $S^{ns}_{2,2,8}$ read
     \begin{eqnarray*}
      S^s_{2,2,8} & = & (\Omega-\gamma) \left[ -\left( \frac{d_r-2c_r\delta}{2} \right) \frac{\left( E_1-D_1\alpha \right)}{\gamma}   
      +\left( \frac{c_r( \delta ^2-\Omega ^2 )-\delta d_r}{2\Omega} \right) D_1 \right]\\
      & \, & + (\Omega -\gamma) \frac{4}{3} \frac{\left( K_1 -2Q_1\alpha \right)}{\left( \alpha^2+\gamma^2\right)^{1/4}} 
      \left[ -\left( \frac{d_r-2c_r\delta}{2} \right) \cos(\phi/2)
      -\left( \frac{c_r( \delta ^2-\Omega ^2 )-\delta d_r}{2\Omega} \right) \sin(\phi/2) \right]  \\
      & \, & + (\Omega-\gamma)\frac{4}{3} \left[ \frac{ -K_1 \alpha +Q_1\left(\alpha^2-\gamma^2 \right) }{\gamma \left( \alpha^2+\gamma^2\right)^{1/4}}\right]
      \left( \frac{d_r-2c_r\delta}{2} \right)\sin(\phi/2) \\
       & \, &  - (\Omega - \gamma)\frac{4}{3} \left[ \frac{ -K_1 \alpha +Q_1\left(\alpha^2-\gamma^2 \right) }{\gamma \left( \alpha^2+\gamma^2\right)^{1/4}}\right]
     \left( \frac{c_r( \delta ^2-\Omega ^2 )-\delta d_r}{2\Omega} \right) \cos(\phi/2) \\
      S^{ns}_{2,2,8} & = & -\frac{\Sto}{\alpha_r}(\Omega-\gamma)\rme^{-(\alpha+\delta)\tau_p}
      \left[ -Z_6\left( \frac{d_r-2c_r\delta}{2} \right) 
      + Z_5\left( \frac{c_r( \delta ^2-\Omega ^2 )-\delta d_r}{2\Omega}\right)  \right]
      \cos\left((\Omega-\gamma)\tau_p\right)  \\
      & \, & - \frac{\Sto}{\alpha_r}(\Omega-\gamma)\rme^{-(\alpha+\delta)\tau_p}
      \left[  -Z_5\left( \frac{d_r-2c_r\delta}{2} \right) 
      - Z_6\left( \frac{c_r( \delta ^2-\Omega ^2 )-\delta d_r}{2\Omega}\right) \right] 
        \sin\left((\Omega-\gamma)\tau_p\right)\\
      & \, & + \frac{(\Omega-\gamma)}{2}\rme^{-\alpha\tau_p}
      \left[- Z_8\cos\left((\Omega-\gamma)\tau_p\right) - Z_7\sin\left((\Omega-\gamma)\tau_p\right) \right]\cdot \\
      & \, & \left[ T_1 \left( \Omega\cos(\Omega\tau_p) + \delta\sin(\Omega\tau_p) \right)
        + T_2 \left( -\delta\cos(\Omega\tau_p) + \Omega\sin(\Omega\tau_p) \right) \right] \\
    & \, & + \frac{(\Omega-\gamma)}{2}\rme^{-\alpha\tau_p}
    \left[ Z_7\cos\left((\Omega-\gamma)\tau_p\right) - Z_8\sin\left((\Omega-\gamma)\tau_p\right) \right]\cdot \\
    & \, & \left[ T_1 \left( \Omega\sin(\Omega\tau_p) - \delta\cos(\Omega\tau_p) \right) 
    + T_2 \left( -\delta\sin(\Omega\tau_p) - \Omega\cos(\Omega\tau_p) \right) \right] 
      \end{eqnarray*}
  %


\bibliographystyle{jfm}

\end{document}